\documentclass[prl,twocolumn,showpacs,amsmath,amssymb,superscriptaddress]{revtex4-2}

\usepackage{amsmath,amssymb,amsfonts,bm,color,graphicx,tabularx}

\usepackage[unicode=true,colorlinks=true]{hyperref}

\usepackage[etex=true,export]{adjustbox}
\usepackage{makecell}
\usepackage{multirow}
\usepackage{amsmath}
\usepackage{verbatim}

\hypersetup{linkcolor=blue, citecolor=blue,urlcolor=blue}
\usepackage{tabularx,multirow,array,diagbox}
\usepackage{adjustbox}

\newcommand{\beq}{\begin{equation}}
\newcommand{\eeq}{\end{equation}}
\newcommand{\beql}{\begin{equation*}}
\newcommand{\eeql}{\end{equation*}}
\newcommand{\beqn}{\begin{eqnarray}}
\newcommand{\eeqn}{\end{eqnarray}}

\usepackage{chemformula} 
\usepackage[T1]{fontenc} 

\usepackage{amsfonts}
\usepackage{mathrsfs}
\usepackage{amsmath}
\usepackage{color}
\usepackage{graphicx}
\graphicspath{ {./pictures/v2/} }
\usepackage{bm}
\usepackage{amssymb}
\usepackage{xspace}
\usepackage{epstopdf}
\usepackage{dcolumn}
\usepackage{longtable}
\usepackage{multirow}
\usepackage{float}
\usepackage{comment}

\usepackage{lipsum}
\usepackage{tabularx}
\usepackage{booktabs}
\def\ie{{\it i.e.},\ }

\usepackage{physics}

\begin{document}
\title{Unconventional Magnetism: Symmetry Classification, Hybrid-parity and Unconstrained-parity Classes}

\author{Xun-Jiang Luo}
\email{xjluo@hmfl.ac.cn}
\affiliation{Anhui Province Key Laboratory of Low-Energy Quantum Materials and Devices, 
High Magnetic Field Laboratory, HFIPS, Chinese Academy of Sciences, Hefei, Anhui 
230031, China }
\author{Dan Li}
\affiliation{School of Physical Science and Technology, Inner Mongolia University, Hohhot 010021, China}
\author{Rui-Chun Xiao}
\affiliation{Institute of Physical Science and Information Technology, Anhui University, Hefei 230601, China}
\author{Ding-Fu Shao}
\affiliation{Key Laboratory of Materials Physics, Institute of Solid State Physics,
HFIPS, Chinese Academy of Sciences, Hefei 230031, China}
\author{Lei Li}
\email{lilei1993@imu.edu.cn}
\affiliation{School of Physical Science and Technology, Inner Mongolia University, Hohhot 010021, China}
\affiliation{Research Center for Quantum Physics and Technologies, Inner Mongolia University, Hohhot 010021, China}
\affiliation{Inner Mongolia Key Laboratory of Microscale Physics and Atomic Manufacturing, Inner Mongolia University, Hohhot 010021, China}
\author{Mingliang Tian}
\email{tianml@hmfl.ac.cn}
\affiliation{Anhui Province Key Laboratory of Low-Energy Quantum Materials and Devices, 
High Magnetic Field Laboratory, HFIPS, Chinese Academy of Sciences, Hefei, Anhui 
230031, China }
\author{Yugui Yao}
\email{ygyao@bit.edu.cn}
\affiliation{Centre for Quantum Physics, Key Laboratory of Advanced Optoelectronic Quantum Architecture and Measurement (MOE), School of Physics, Beijing Institute of Technology, Beijing, 100081, China }

\begin{abstract}

Unconventional magnetism has emerged as a transformative frontier in condensed matter physics. Such phases are characterized by substantial non-relativistic spin splitting (NSS) in symmetry-compensated magnets. They have been classified by the parity of their spin textures under momentum inversion, leading to the paradigms of altermagnets (even-parity) and odd-parity magnets. However, the full symmetry landscape remains largely unexplored. In this Letter, we present a systematic classification framework for unconventional magnetism based on the representation theory of the spin textures and the associated parity properties. Within this framework, we predict two previously unidentified classes beyond the established pure-parity categories: hybrid-parity magnets (HPMs) and unconstrained-parity magnets (UPMs), where the spin textures exhibit contrasting parities among their Cartesian components and the parity of the spin textures is ill-defined, respectively. We derive universal symmetry criteria that categorize HPMs into three distinct types. Importantly, by combining the spin splitting characteristics of altermagnets and odd-parity magnets, HPMs can enable the coexistence of the spin current and Edelstein effects. Taking FePO$_4$ as an example, we perform first-principles calculations to demonstrate this coexistence. Finally, we discuss the potential applications of HPMs in spintronic devices. Our work provides a comprehensive symmetry classification of unconventional magnetism and establishes HPMs as a promising platform for multi-functional spintronics.
\end{abstract}

\maketitle

\textit{Introduction.}---Unconventional magnetism (UM) has emerged as a pivotal frontier in condensed matter physics \cite{Hayami2019,Yuan2020a,Libor2020,Ma2021,Feng2022,Tomas2022,ifmmode2022a,shaodingfu2023,Krempaský2024,ZhangRunWu2024,ZhouXiaodong2024,Reimers2024,BaiLing2024,McClarty2024,LeeSuyoung2024,Shao2024,PhysRevLett.134.026001,Chen2025,Zhang2025a,Liu2025a,2026arXiv260327505C,2026arXiv260421397Y}, transcending the traditional dichotomy between ferromagnetic and antiferromagnetic paradigms. These systems are distinguished by substantial non-relativistic spin splitting (NSS) in the presence of symmetry-compensated magnetization (SCM) \cite{Liu2025a,2026arXiv260327505C,2026arXiv260421397Y}. This unique synergy merges the ultrafast dynamics of antiferromagnets with the spintronic functionalities of ferromagnets, opening new avenues for next-generation spintronics, such as magnetic tunnel junctions \cite{Shao2021}, spin-splitter torque devices \cite{songchneg2023,LeiHan2024,Gon2021}, and spin transistors \cite{j3qj-77yj}. The recently developed spin space group (SSG), which decouples symmetry operations in lattice and spin spaces, provides a rigorous symmetry framework for describing such magnetic phases \cite{LiuPengfei2022, CheXiaobing2024, JiangYi2024, XiaoZhenyu2024, HanaSchiff2025}. In particular, based on the parity of NSS under momentum inversion, UM has been categorized into two fundamental classes: altermagnets (even-parity) and odd-parity magnets \cite{ifmmode2022, HayamiSatoru2020, BirkHellenes2023, 2026Luoxun}, which exhibit symmetric or antisymmetric spin textures in momentum space, respectively. Despite significant progress in the symmetry classification of UM \cite{HuMengli2025,Yuntian2025,2025song,2026Luoxun,xiaobingchen2026}, a complete taxonomy
remains an open question.

\begin{figure}
\centering
\includegraphics[width=3.5in]{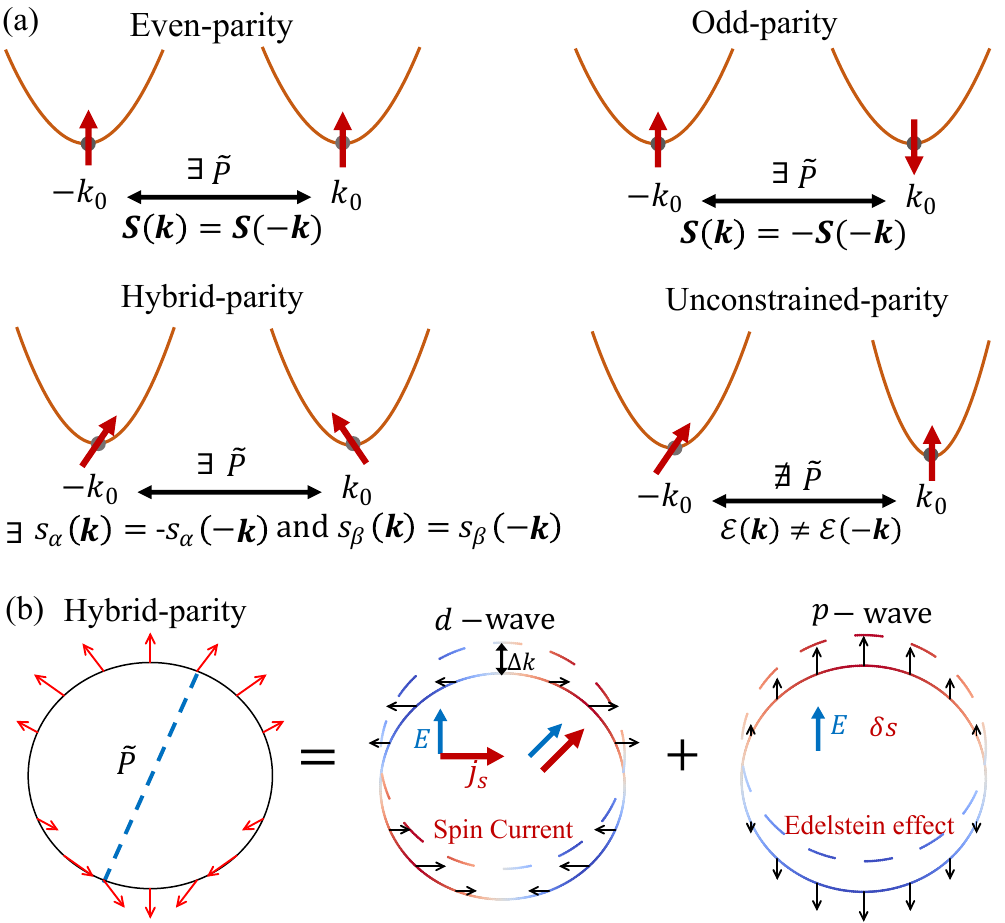}
\caption{(a) Schematic illustration of the even‑parity, odd‑parity, hybrid‑parity, and unconstrained‑parity classes of UM. $\tilde{P}$ denotes the SSG symmetry that flips momentum. $\bm{S}(\bm{k}) = (s_x(\bm{k}), s_y(\bm{k}), s_z(\bm{k}))$ is the spin expectation value. $\mathcal{E}(\bm{k})$ is the energy at momentum $\bm{k}$. (b) Illustration of a hybrid‑parity spin texture and its decomposition into $d$-wave and $p$-wave channels along different spin directions, thereby leading to the spin current ($j_s$) and Edelstein effect under an applied electric field $\bm E$.}
\label{Fig1}
\end{figure}

\begin{figure*}
\centering
\includegraphics[width=7in]{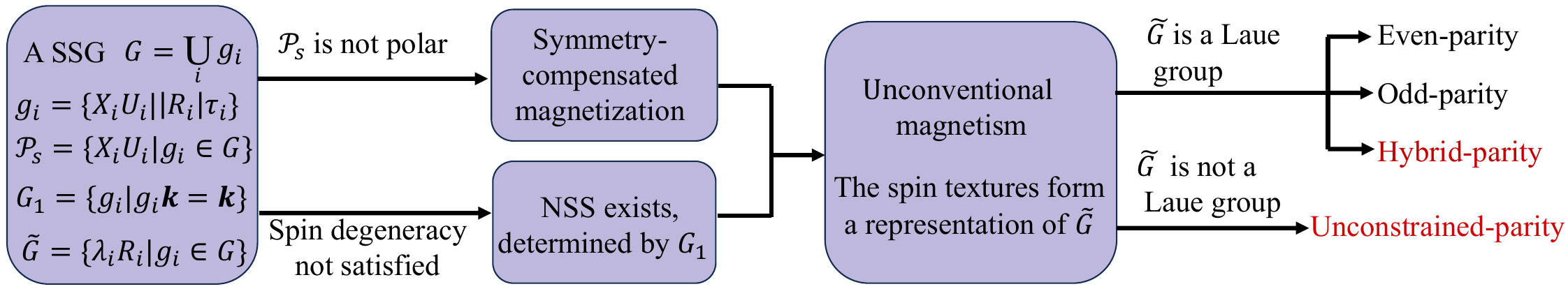}
\caption{Flowchart illustrating the symmetry classification of UM. Starting from the SSG $G$, the derived point group $\mathcal{P}_s$ and $G_1$ dictate the conditions for SCM and NSS, respectively. The spin textures form a representation of the  point group $\tilde{G}$. If $\tilde{G}$ is a Laue group, the spin texture acquires a definite parity, leading to even-, odd-, or hybrid-parity classes; if not, the unconstrained-parity class arises. Here $\lambda_i = \det(X_iU_i)$.}
\label{Fig2}
\end{figure*}

In this Letter, we present a unified classification framework for UM based on the SSG theory. Within this framework, we predict two previously unidentified classes beyond the established even- and odd-parity magnets: hybrid-parity magnets (HPMs) and unconstrained-parity magnets (UPMs). In HPMs, different Cartesian components of the spin texture exhibit mixed parities (some even, some odd), whereas UPMs lack a well-defined parity. Figure.~\ref{Fig1}(a) schematically depicts these four distinct classes of UM. The classification scheme is summarized in Fig.~\ref{Fig2}. Starting from a given SSG $G$, we determine the symmetry conditions for UM, namely, the requirements for SCM and NSS. In particular, momentum-dependent spin textures generally transform as a representation of the point group $\widetilde{G}$ derived from $G$ \cite{XiaoZhenyu2024}. When $\widetilde{G}$ is a Laue group (momenta $\bm{k}$ and $-\bm{k}$ are related by an SSG symmetry), each component of the spin textures carries a definite parity  (even or odd) under momentum inversion. This regime encompasses the known even- and odd-parity magnets \cite{2026Luoxun}, as well as the newly predicted HPMs. Conversely, when $\widetilde{G}$ is not a Laue group, $\bm{k}$ and $-\bm{k}$ are not symmetry-related, rendering the parity ill-defined and thereby defining the UPMs.

In a preceding work by one of us (Ref.~\cite{2026Luoxun}), the symmetry criteria of even- and odd-parity magnets have been derived. Here, we develop a comprehensive symmetry framework that unifies the established even- and odd-parity classes with the newly identified HPMs and UPMs. We derive universal symmetry criteria that classify HPMs into three distinct types according to their spin-texture characteristics. Applying these criteria, we identify material candidates for both HPMs and UPMs from the MAGNDATA database. Crucially, by combining the spin-splitting characteristics of even- and odd-parity magnets, HPMs can enable the coexistence of spin current and Edelstein responses, as schematically illustrated in Fig.~\ref{Fig1}(b). Taking FePO$_4$ as a prototypical example, we perform first-principles calculations to demonstrate its distinctive spin-splitting features and study the resulting spin transport properties. Finally, we discuss the potential applications of HPMs in spintronic devices.

\textit{Classification framework.}---A general SSG operation is denoted as \( g_i = \{X_i U_i \| R_i | \tau_i\} \), where $\{R_i | \bm{\tau}_i\}$ acts on real space (with $R_i$ being a point group operation and $\bm{\tau}_i$ being translation or zero vector), $U_i$ denotes the proper spin rotation, and \(X_i\) is either the identity operator \(I\) or the time-reversal operator \(T\). To derive the symmetry conditions for UM, we introduce two point groups derived from the full SSG $G$:
\begin{align}
    \mathcal{P}_s &= \{X_i U_i \mid g_i \in G\}, \nonumber\\
    G_1 &= \{g_i \mid g_i \in G,\; g_i \bm{k} = \bm{k}\}.
\end{align}
 A nonpolar $\mathcal{P}_s$ implies zero magnetization enforced by SSG symmetry \cite{Yuntian2025}, a prerequisite for UM.  $G_1$ is the little group at a generic momentum $\bm{k}$ and governs the local properties of NSS \cite{XiaoZhenyu2024}. Specifically, NSS emerges when $G_1$ contains neither $PT$ symmetry nor multiple spin-rotation-translation symmetries with distinct rotation axes \cite{LiuPengfei2022}. When NSS exists, $G_1$ further determines the dimensionality ($d_s$, the number of non-zero components) of the spin texture $\bm{S}(\bm{k}) = (s_x(\bm{k}), s_y(\bm{k}), s_z(\bm{k}))$, which is defined as the expectation value of the spin operator in the Bloch state. Depending on the symmetries present in $G_1$, the spin textures are classified as follows: collinear ($d_s=1$) textures are guaranteed by $g_1 = \{ U_{\bm n}(\theta) \| I | \bm{\tau}_1 \}$ (aligning spins along $\bm n$); coplanar ($d_s=2$) textures are guaranteed by $g_2 = \{ T U_{\bm n}(\pi) \| P | \bm{\tau}_2 \}$ (confining spins to the plane perpendicular to $\bm n$); noncoplanar ($d_s=3$) textures occur when neither $g_1$ nor $g_2$ is present. Here, $U_{\bm n}(\theta)$ denotes a spin rotation by angle $\theta$ about the $\bm n$-axis. When $\mathcal{P}_s$ is polar and NSS exists (as assumed in the rest of this work), an UM is realized.

Before classifying UM, we first consider the transformation of 
$\bm{S}(\bm{k})$ under the symmetry $g_i = \{X_i U_i \| R_i | \bm{\tau}_i\}$:
\begin{equation}
    g_i\bm{S}(\bm{k}) = U_i \bm{S}(\lambda_i R_i^{-1}\bm{k}),
    \label{eq:constraint}
\end{equation}
where $\lambda_i = \det(X_iU_i) = \pm 1$ \cite{JiangYi2024}.  Importantly, under the full symmetry constraints of the SSG $G$, the non-zero components of $\bm{S}(\bm{k})$ constitute a $d_s$-dimensional representation of the point group $\widetilde{G} = \{\lambda_i R_i \mid g_i \in G\}$ acting on $\bm{k}$ \cite{XiaoZhenyu2024}. The representation matrix is obtained by restricting the spin rotation $U_i$ to the subspace spanned by these non-zero components, as detailed in the Supplemental Material (SM) \cite{supp}. This representation rigorously captures the symmetry properties of $\bm{S}(\bm{k})$, and its Cartesian basis functions determine the explicit form of NSS (see the SM \cite{supp}).

UM has been typically classified by the parity of their spin textures under momentum inversion, which implicitly requires $\widetilde{G}$ to be a Laue group containing the inversion $\widetilde{P}$ (corresponding to $\lambda_i R_i \bm{k} = -\bm{k}$). This effective inversion originates from four distinct symmetry operations: $g_3 = \{ T U_{3} \| I|\bm{\tau}_3 \}$, $g_4 = \{ U_{4} \| P|\bm{\tau}_4 \}$, $g_5 = \{ T \| I |\bm{\tau}_5 \}$, and $g_6 = \{ I \| P|\bm{\tau}_6 \}$, where $g_{3,4,5,6}\bm{k} = -\bm{k}$ and $P$ denotes the spatial inversion. 
Under $\widetilde{P}$, each non-zero component $s_{\alpha}(\bm{k})$ ($\alpha\in{x,y,z}$) acquires a definite parity (see the SMs \cite{supp}):
\begin{align}
   \tilde{P} s_{\alpha}(\bm{k}) = p_{\alpha} s_{\alpha}(-\bm{k}),
\end{align}
where $p_{\alpha} = \pm 1$. This yields three fundamental classes:
\begin{align}
    &\text{even-parity}: \quad p_{\alpha} = 1 \text{ for all } \alpha \text{ with } s_{\alpha} \neq 0; \nonumber \\
    &\text{odd-parity}: \quad p_{\alpha} = -1 \text{ for all } \alpha \text{ with } s_{\alpha} \neq 0; \nonumber \\
    &\text{hybrid-parity}: \quad \exists\, p_{\alpha} \neq p_{\alpha'} \text{ for } s_{\alpha,\alpha'} \neq 0 ,
\end{align}
where $\alpha\neq \alpha^{\prime}$ with $\alpha^{\prime}\in {x,y,z}$. Collinear spin textures can only realize even- and odd-parity classes, which have been extensively studied \cite{DuanXunkai2025,GuMingqiang2025,Song2025a,Yamada2025a}. For coplanar and noncoplanar spin textures, the hybrid-parity class can naturally emerge in addition to pure-parity configurations, representing a previously unexplored regime of UM.

When $\widetilde{G}$ lacks inversion symmetry $\widetilde{P}$ (i.e., none of the symmetries $g_3$--$g_6$ is present in $G$), the parity of the spin texture becomes ill-defined. In this regime, we identify a distinct class of unconventional magnetism, which we term unconstrained-parity magnets (UPMs). These systems are characterized by the coexistence of NSS and SCM, with the defining feature that no SSG symmetry relates $\bm{k}$ and $-\bm{k}$. This scenario occurs exclusively in noncoplanar magnetic orders \cite{Hu2025}, since collinear and coplanar configurations typically preserve $g_3$. Here, we emphasize two points concerning UPMs. First, because the parity of $s_{\alpha}(\bm{k})$ is not constrained, it can be expanded as a sum of even and odd functions of $\bm{k}$, resulting in mixed-parity behavior for each nonzero component of $\bm{S}(\bm{k})$. Second, the energy bands of UPMs are expected to exhibit an asymmetric structure since no symmetry relates opposite momenta, positioning them as a unique platform for studying nonreciprocal transport \cite{annurev}.


\textit{Symmetry criteria for HPMs}---
We now establish the symmetry criteria for HPMs and present their classification. Since the definition of HPMs involves only the constraints between \( \bm S(\bm k) \) and \( \bm S(-\bm k) \), their symmetry criteria are fully determined by the symmetries that flip or preserve the momentum. These include \( g_1 - g_6 \) with \( g_{1,2}\bm k = \bm k \) and \( g_{3,4,5,6}\bm k = -\bm k \). However, \( g_5 \) and \( g_6 \) enforce pure odd- and even-parity, respectively, while \( g_1 \) forces the spin texture to be collinear. Consequently, only symmetries \( g_2 \), \( g_3 \), and \( g_4 \) are compatible with HPMs.


The specific constraints imposed by $g_{2}$, $g_{3}$, and $g_{4}$ on  $\bm S(\bm k)$ depend on the form of the spin rotation $U_i$. For our purpose, we take $U_{2} =U_{y}(\pi)$, $U_{3} = U_{x}(\pi)$, and $U_{4} = U_{z}(\pi)$ \cite{supp}. Under these choices, $g_2$-$g_4$ generate the symmetry constraints
\begin{align}
g_{2}:\quad \bm{S}(\bm{k}) &= \big( s_x(\bm{k}), 0, s_z(\bm{k}) \big), \nonumber \\
g_{3}:\quad \bm{S}(\bm{k}) &= \big( -s_x(-\bm{k}), s_y(-\bm{k}), s_z(-\bm{k}) \big), \nonumber \\
g_{4}:\quad \bm{S}(\bm{k}) &= \big( -s_x(-\bm{k}),-s_y(-\bm{k}),\, s_z(-\bm{k}) \big).
\label{eq4}
\end{align}
Under these symmetry constraints, we classify HPMs into the following three types:

\begin{itemize}
    \item \textbf{Type-I (Coplanar hybrid-parity):} The spin texture lies in a plane and satisfies $s_{\alpha}(\bm{k}) = -s_{\alpha}(-\bm{k})$, $s_{\beta}(\bm{k}) = s_{\beta}(-\bm{k})$, and $s_{\gamma}(\bm{k}) = 0$ with $\alpha \neq \beta \neq \gamma$. This type is realized by the combination of $g_{3}$ and $g_{4}$, leading to $s_z(\bm{k}) = s_z(-\bm{k})$ and $s_x(\bm{k}) = -s_x(-\bm{k})$ (with $s_y(\bm{k}) = 0$). Note that combining $g_2$ with $g_3$ or $g_4$ yields the same constraints, but these combinations are not independent because $g_4=g_2g_3$ and $g_3=g_2g_4$.

    \item \textbf{Type-II (Noncoplanar with two even components):} The spin texture is noncoplanar and obeys $s_{\alpha,\beta}(\bm{k}) = s_{\alpha,\beta}(-\bm{k})$ and $s_{\gamma}(\bm{k}) = -s_{\gamma}(-\bm{k})$. It is realized by $g_{3}$ alone, giving $s_{y,z}(\bm{k}) = s_{y,z}(-\bm{k})$, $s_{x}(\bm{k}) = -s_{x}(-\bm{k})$.

    \item \textbf{Type-III (Noncoplanar with two odd components):} This type also features a noncoplanar spin texture, satisfying  $s_{\alpha,\beta}(\bm{k}) = -s_{\alpha,\beta}(-\bm{k})$ and $s_{\gamma}(\bm{k}) = s_{\gamma}(-\bm{k})$. This type is realized by $g_{4}$ alone, yielding $s_{z}(\bm{k}) = s_{z}(-\bm{k})$, $s_{x,y}(\bm{k}) = -s_{x,y}(-\bm{k})$.
    
\end{itemize}

We discuss the realization of HPMs in different magnetic orders. Collinear magnetic orders host only collinear spin textures, and therefore cannot support HPMs. In coplanar magnetic orders (e.g., lying in the $yz$-plane), the symmetry $g_3=\{TU_x(\pi)||I|\bm 0\}$ is typically preserved. If the additional symmetry $g_4$ is also present, this realizes Type-I HPMs. Otherwise, if only $g_3$ is present, Type-II HPMs are realized. In noncoplanar magnetic orders, the symmetry $g_3$ is generally broken and therefore only Type-III HPMs can be realized under the symmetry $g_4$ alone.

Importantly, the above classification framework based on the presence or absence of symmetries $g_1$-$g_6$ is general for diagnosing the parity properties of spin textures, including even-, odd-, hybrid-, and unconstrained cases.  In the SMs \cite{supp}, we construct theoretical models with distinct magnetic moments on the honeycomb lattice to illustrate this unified symmetry principle. Notably, the mechanism guaranteeing SCM differs between odd-parity magnets and the other three classes. For odd-parity NSS, SCM is inherently satisfied because the preserved symmetries of $g_1$-$g_6$ enforce opposite spins at $\bm{k}$ and $-\bm{k}$. By contrast, even-parity magnets, HPMs, and UCMs require additional symmetries satisfying $g_i\bm{k} \neq \pm \bm{k}$ to enforce zero net magnetization, supplementing the preserved symmetries of $g_1$-$g_6$.

\begin{figure}
\centering
\includegraphics[width=3.5in]{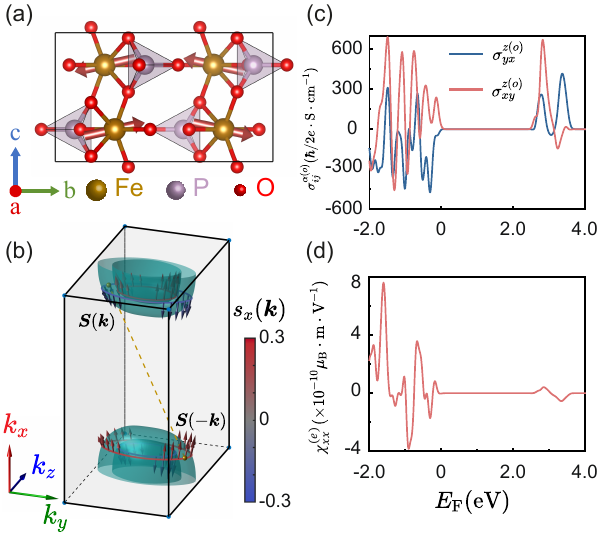}
\caption{(a) Crystal structure and magnetic order of bulk FePO$_4$.  (b) Fermi surface of FePO$_4$ at $E_F = -35$~meV. The spin textures on the $k_x = \pm 0.5$~Å$^{-1}$ planes are plotted; arrows indicate the orientation of vector $\bm S(\bm k)$, and the color encodes the out-of-plane spin component $s_x(\bm{k})$. (c) Calculated $T$-odd spin Hall conductivity $\sigma_{xy}^{z(o)}$ and $\sigma_{yx}^{z(o)}$ as a function of the Fermi energy $E_F$. (d) $T$-even Edelstein susceptibility $\chi_{xx}^{(e)}$ as a function of $E_F$. }
\label{Fig3}
\end{figure}

\textit{Candidate materials and spin transport}---
By applying the symmetry criteria for parity of $\bm S(\bm k)$ and examining the SCM condition, we identify 29 HPM and 20 UPM candidate materials from the MAGNDATA database using the online program FINDSPINGROUP~\cite{CheXiaobing2024}. Among the HPM candidates, 11 are of type-I, 17 of type-II, and 1 of type-III. Among the 20 UPM candidates, CrSe and TbCrO$_3$ exhibit collinear spin textures, whereas the remaining 18 materials exhibit noncoplanar spin textures.



The Fermi-surface-mediated spin transport properties are intimately linked to the spin texture $\bm{S}(\bm{k})$ at the Fermi level. Specifically, even-parity spin textures can generate spin currents \cite{Binghai2017,Kimata2019,Naka2019}, whereas odd-parity textures can give rise to the Edelstein effect \cite{EDELSTEIN1990233}. These two phenomena have been extensively studied in altermagnets and odd-parity magnets, respectively \cite{Shao2021,Gon2021,Bose2022,bf1n-sxdl2025,Chakraborty2025}. In HPMs, different spin components exhibit contrasting parities, as schematically depicted in Fig.~\ref{Fig1}(b). This texture can be decomposed into $d$-wave and $p$-wave channels along orthogonal spin directions. The $d$-wave channel yields a spin-current response whose geometric relation between the current flow and the electric field is identical as in $d$-wave altermagnets \cite{Gon2021},  while the $p$-wave channel gives rise to the Edelstein effect under an applied electric field $E$. These responses are described by the tensors $\sigma_{ij}^{\alpha (o)}$ and $\chi_{\alpha j}^{(e)}$, defined through $J_{i}^{\alpha} = \sigma_{ij}^{\alpha (o)} E_j$ and $\delta s_{\alpha} = \chi_{\alpha j}^{(e)} E_j$, where $\alpha$ labels the spin polarization direction, $i$ the spin-current direction, and $j$ the electric-field direction. The superscripts $(o)$ and $(e)$ distinguish the time-reversal-odd and time-reversal-even Fermi-surface responses, respectively \cite{supp}. In the following, we take the type-I HPM FePO$_4$ as a representative example to demonstrate the hybrid-parity spin texture and the resulting spin transport properties.


FePO$_4$ hosts a coplanar magnetic order (N\'{e}el temperature $T_N = 125$ K \cite{rousse2003magnetic}) with four Fe atoms per unit cell, as shown in Fig.~\ref{Fig3}(a). It is described by the oriented SSG $P^{2_{010}} a^{2_{100}} n^1 m^{m_{100}} 1$ \cite{Yuntian2025}, 
preserving the symmetries $g_3=\{TU_x(\pi)\|I|\bm{0}\}$ and $g_4=\{U_z(\pi)\|P|\bm{0}\}$.
The two symmetries enforce a hybrid-parity spin texture satisfying $s_x(\bm{k}) = -s_x(-\bm{k})$, $s_z(\bm{k}) = s_z(-\bm{k})$, and $s_y(\bm{k}) = 0$. The pair $(s_x(\bm{k}), s_z(\bm{k}))$ forms a reducible two-dimensional representation $B_{1u}\oplus B_{2g}$ of the point group $D_{2h}$. This representation captures the full symmetry properties of the spin textures, which are characterized by $s_x(\bm{k})\propto k_x$ and $s_z(\bm{k})\propto k_x k_y$ (see the SMs \cite{supp}), exhibiting a $p$-wave and $d$-wave pattern, respectively. FePO$_4$ is a semiconductor; its first-principles band structure is provided in the SM \cite{supp}. By shifting the Fermi level into the valence band, we obtain the Fermi-surface spin textures shown in Fig.~\ref{Fig3}(b), consistent with the symmetry analysis.

The symmetry generators of FePO$_4$ are $\{U_{z}(\pi) \| P|\bm 0\}$, $\{ U_{y}(\pi) \| M_x|\bm{\tau}_{(1/2,1/2,0)}\}$, $\{I \| M_z|\bm{\tau}_{(0,0,1/2)}\}$, and $\{T U_x(\pi) \| I|\bm 0\}$, with $M_{x}$ and $M_{z}$ being spatial mirrors. These generators impose the following constraints on the response tensors:
\begin{align}
& \sigma_{ij}^{x(o)} = 0,\quad \sigma_{ij}^{y(o)} = 0, \nonumber\\
&\sigma^{z(o)} = \begin{pmatrix}
0 & \sigma_{xy}^{z}  & 0\\
\sigma_{yx}^{z} & 0 & 0 \\
0 & 0 & 0
\end{pmatrix},\quad
\chi^{(e)} = \begin{pmatrix}
\chi_{xx} & 0 & 0 \\
0 & 0 & 0 \\
0 & 0 & 0
\end{pmatrix}.
\end{align} 
In Figs.~\ref{Fig3}(c) and \ref{Fig3}(d), we plot the nonzero elements of $\sigma^{z(o)}$ and $\chi^{(e)}$ as functions of the chemical potential, demonstrating the coexistence of the magnetic spin Hall and Edelstein effects. 
Detailed band structures, transport calculations, and spin-orbit coupling effects of FePO$_4$ are provided in the SMs \cite{supp}.

We note that the spin current and Edelstein responses in FePO$_4$ are distinct from those previously reported in altermagnets (e.g., RuO$_2$ \cite{Gon2021}) and $p$-wave magnets (e.g., CeNiAsO \cite{Chakraborty2025}). Although FePO$_4$ hosts a coplanar spin texture, its spin Hall and Edelstein responses each involve only a single spin-polarization component. This collinear nature of the spin current ensures long-distance spin transport, while the collinear odd-parity spin splitting enables high-efficiency charge-spin conversion, similar to that in $p$-wave magnetic systems \cite{Chakraborty2025}. In contrast, CeNiAsO and RuO$_2$ exhibit collinear spin textures that permit only the Edelstein effect and the spin current (with spin polarization along the N\'{e}el vector direction) response, respectively. This dichotomy is rigorously symmetry-enforced: as demonstrated in the SMs \cite{supp}, the preserved spin-only symmetry $\infty m$ in altermagnets enforces $\chi_{\alpha j}^{(e)}=0$; in odd-parity magnets, the symmetries that enforce the odd-parity spin splitting also enforce $\sigma_{ij}^{\alpha(o)}=0$. HPMs evade both restrictions, enabling the coexistence of spin current and Edelstein responses. Moreover, although we use type-I HPM as an example, this coexistence is a general phenomenon across all HPM types (type-I to type-III). In particular, type-II HPMs can enable the generation of spin currents with multidirectional spin polarizations \cite{supp}.

\begin{figure}
\centering
\includegraphics[width=3.5in]{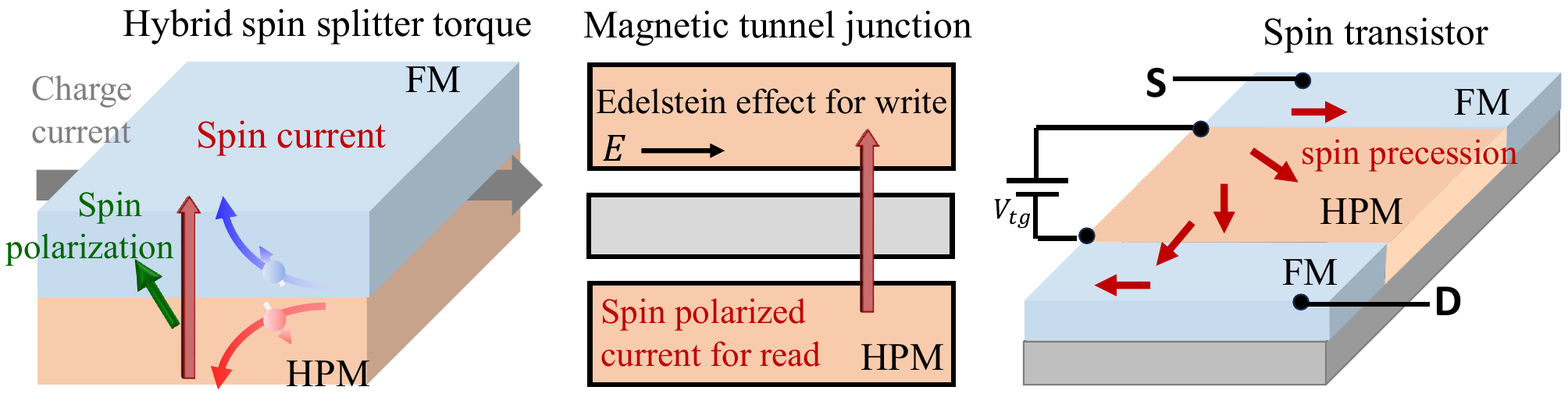}
\caption{Proposed spintronic device architectures enabled by HPMs. (a) Hybrid spin-splitter torque: a charge current injected into an HPM/FM bilayer generates spin currents with multi spin polarizations components. (b) Magnetic tunnel junction: the top layer utilizes the Edelstein effect for electrical write operations, while the bottom HPM layer provides a spin-polarized readout current. (c) Spin transistor: an HPM channel sandwiched between ferromagnetic (FM) source (S) and drain (D) contacts; gate voltage $V_{tg}$ modulates spin precession in the HPM, enabling non-volatile logic functionality.} 
\label{Fig4}
\end{figure}

\textit{Conclusion and Discussion}---In summary, we establish a unified symmetry framework for classifying UM based on their parity properties. Our framework predicts two previously unidentified classes: HPMs and UPMs. We classify HPMs into three distinct types and derive their symmetry criteria. Importantly, we demonstrate that the unique hybrid-parity spin textures can give rise to multifunctional spin transport phenomena.

Based on the novel spin transport properties, three prototypical spintronic devices could be realized with HPMs, as illustrated in Fig.~\ref{Fig4}. In a hybrid spin-splitter torque device, a charge current injected into an HPM/Ferromagnet (FM) bilayer generates spin currents carrying multiple spin-polarization components, enabling field-free spin torque switching \cite{RevModPhys.91.035004}. A magnetic tunnel junction comprising two HPM layers separated by a thin tunnel barrier exploits the Edelstein effect in the top layer for electrical write operations, while the bottom layer provides a spin-polarized readout current, achieving a compact all-HPM read-write stack. Furthermore, a spin transistor with an HPM channel sandwiched between ferromagnetic source and drain contacts allows gate voltage $V_{tg}$ to modulate spin precession in the HPM, enabling non-volatile logic functionality. These architectures leverage the hybrid-parity nature of HPMs to integrate multiple spin functionalities within a single material platform.

We emphasize that HPMs inherit the spin-splitting characteristics of both altermagnets and odd-parity magnets, but generally exhibit lower SSG symmetries than either class. The reduced symmetry lifts additional constraints, thereby permitting a broader range of physical phenomena. For example, such materials may host the anomalous Hall effect \cite{Yuntian2025}, the nonlinear Hall effect \cite{Zhu2025}, and (spin) photovoltaic effects \cite{463y-q7lt,PhysRevLett.134.196907,2025arXiv251217315P}. Moreover, some predicted HPM candidates, including  YMnO$_3$ and LuMnO$_3$, are known multiferroics \cite{supp,Dong02112015}. The coexistence of ferroelectricity and hybrid-parity spin textures in these compounds suggests the possibility of electrically manipulating the spin splitting \cite{DuanXunkai2025,GuMingqiang2025}, paving the way for electrically controlled spintronics devices \cite{2026arXiv260101499Y}. Thus, our work establishes HPMs as a promising platform that may host rich phenomena, encompassing spin-transport responses, unconventional superconductivity \cite{PhysRevB.108.184505,Zhangsongbo2024,52wh-1z5y,2026arXiv260315147L}, and other desirable functionalities inherited from both altermagnets and odd-parity magnets.

\textit{Note added.} We note that a recent work proposes using Floquet engineering to break the symmetry that relates $\bm{k}$ and $-\bm{k}$ in collinear magnetic orders \cite{zhengyang2026}. The resulting systems still exhibit both SCM and NSS, thus falling into the class of UPMs in our framework.

\section{Acknowledgment}
Xun-Jiang Luo thanks helpful discussions with Xilin Feng and Mengli Hu. 
L. Li is supported by the National Natural Science Foundation of China (Grant No. 12504107) and the Startup Project of Inner Mongolia University (Grant No. 10000-A260015/301). 
R-C. X. is supported by the National Natural Science Foundation of China (Grant No. 12474100)
Y.Y. is supported by the NSF of China (Grants Nos. 12321004, W2511003, 12234003).

\bibliography{reference}

\clearpage

\begin{widetext}
\begin{center}
\begin{large}
\textbf{Supplemental Material for ‘‘Unconventional Magnetism: Symmetry Classification, Hybrid-parity and
Unconstrained-parity classes"}
\end{large}
\end{center}

\setcounter{figure}{0}
\setcounter{equation}{0}
\renewcommand\thefigure{S\arabic{figure}}
\renewcommand\thetable{S\arabic{table}}
\renewcommand\theequation{S\arabic{equation}}

This Supplemental Material includes the following nine sections:
(1) Symmetry properties of the spin textures and application to FePO$_4$;
(2) The parity of the spin textures;
(3) A unified symmetry classification of even-, odd-, and hybrid-parity magnets;
(4) Candidate materials for hybrid- and unconstrained-parity magnets;
(5) Spin transport response formalism;
(6) Theoretical models;
(7) First-principle calculation of FePO$_4$;
(8) Response tensors in altermagnets and odd-parity magnets;
(9) Type-II and type-III HPMs.

\section{A. Symmetry properties of the spin textures and application to FePO$_4$}
In this section, we explain why the nonzero components of the spin texture $\bm{S}(\bm{k})$ form a representation of the point group $\tilde{G}$ introduced in the main text, and how this representation determines the allowed functional form of nonzero spin expectation values $s_{\alpha}(\bm{k})$ ($\alpha = x, y, z$).

Under a spin space group (SSG) operation $g_i = \{X_iU_i \| R_i | \bm{\tau}_i\}$, the spin texture transforms as
\begin{equation}
g_i \bm{S}(\bm{k}) = U_i \bm{S}(\lambda_i R_i^{-1}\bm{k}), \qquad \lambda_i = \det(X_iU_i) = \pm 1,
\label{eq:app_trans}
\end{equation}
 where $\{R_i | \bm{\tau}_i\}$ acts on real space (with $R_i$ being a point group operation and $\bm{\tau}_i$ being a translation or zero vector), $U_i$ denotes the proper spin rotation, and \(X_i\) is either the identity operator \(I\) or the time-reversal operator \(T\). Each SSG operation $g_i$ can be mapped to a pair $(U_i, \lambda_i R_i)$. The spin texture $\bm{S}(\bm{k})$ transforms under the matrices $U_i$, thereby forming a representation of the SSG $G = \{g_i\}$. However, our primary interest lies in the symmetry of $\bm{S}(\bm{k})$ as a function of $\bm{k}$ throughout the Brillouin zone. To capture the dependence of momentum, we focus on the  point group $\widetilde{G} = \{\lambda_i R_i \mid g_i \in G\}$, which acts on the momentum variable $\bm{k}$.
It should be noted that the matrices $\{U_i\}$ do not necessarily form a representation of $\widetilde{G}$ directly. The reason is that different SSG operations may map to the same element $\tilde{g} = \lambda_i R_i \in \widetilde{G}$ but possess different spin-space matrices $U_i$. For example, both time-reversal symmetry $T$ and space inversion symmetry $P$ map to $\tilde{g}=-I$, while the spin operations are distinct. 

More generally, we suppose that two operations $g_a$ and $g_b$ both satisfy $\lambda_a R_a = \lambda_b R_b = \tilde{g}$, and denote their spin matrices as $U_a$ and $U_b$, respectively, with $U_a\neq U_b$. Then, we have 
\begin{align}
g_a:&\quad \bm{S}(\tilde{g}\,\bm{k}) = U_a \bm{S}(\bm{k}), \nonumber\\
g_b:&\quad \bm{S}(\tilde{g}\,\bm{k}) = U_b \bm{S}(\bm{k}).
\end{align}
Subtracting the two equations gives $(U_a - U_b) \bm{S}(\bm{k}) = 0$ for all $\bm{k}$. Thus, $\bm{S}(\bm{k})$ must lie entirely in the null space (kernel) of $(U_a - U_b)$. In this kernel space, the remaining components transform under $\widetilde{G}$ via a well-defined representation matrix $U_{\tilde{g}}$. For example, consider an SSG containing $g_a=\{I\|P|\bm 0\}$ and $g_b= \{T U_z(\pi)\|I|\bm 0\}$ (with $U_z$ being spin rotation around the $z$-axis), both mapping to $\tilde{g} = -I$ in $\widetilde{G}$. Their spin-space matrices are $ U_a=I$ and $ U_b=\operatorname{diag}(-1,-1,1)$, respectively. The symmetry constraints on $\bm S(\bm k)$ lead to $s_x(\bm{k}) = s_y(\bm{k}) = 0$, allowing only $s_z(\bm{k})$ to be non-zero. This surviving component $s_z$ then forms a one-dimensional representation of $\widetilde{G}$ with  $U_{\tilde{g}}=1$. Therefore, the non-zero components of $\bm{S}(\bm{k})$ constitute a $d_s$-dimensional representation of $\widetilde{G}$, where $d_s$ is the number of non-zero components. The representation matrix can be obtained by  restricting the corresponding spin rotation matrix $U_i$ to the subspace spanned by the non-zero components.


Near the $\Gamma$ point ($\bm{k}=0$), each non-zero spin component $s_\alpha(\bm{k})$ (with $\alpha=x,y,z$) can be expanded as a Taylor series:
\begin{equation}
s_\alpha(\bm{k}) = \sum_{i,j,k \ge 0} a_{ijk}^{(\alpha)} \, k_x^i k_y^j k_z^k.
\label{eq:taylor}
\end{equation}
In the following, we show that the nonzero coefficients $a_{ijk}^{(\alpha)}$ are directly related to the Cartesian basis functions of the representation $\rho_s$ of $\widetilde{G}$ formed by $\bm{S}(\bm{k})$ by taking FePO$_4$ as an example.

FePO$_4$ is described by the oriented SSG $P^{2_{010}} a^{2_{100}} n^1 m^{m_{100}} 1$. The symmetry generators are 
\beqn
\{U_{z}(\pi) \| P|\bm 0\},\quad 
\{ U_{y}(\pi) \| M_x|\bm{\tau}_{(1/2,1/2,0)}\},\quad 
\{I \| M_z|\bm{\tau}_{(0,0,1/2)}\},\quad 
\{T U_x(\pi) \| I|\bm 0\},
\eeqn
which generate the point group $\widetilde{G} = D_{2h}$. These generators impose the following constraints on the spin texture $\bm{S}(\bm{k})$:
\begin{align}
\{U_{z}(\pi) \| P\}&:\quad s_{x,y}(\bm{k}) = -s_{x,y}(-\bm{k}),\; s_{z}(\bm{k}) = s_{z}(-\bm{k}), \nonumber\\
\{U_{y}(\pi) \| M_x|\bm{\tau}_{(1/2,1/2,0)}\}&:\quad s_{x,z}(k_x,k_y,k_z) = -s_{x,z}(-k_x,k_y,k_z),\; s_{y}(k_x,k_y,k_z) = s_{y}(-k_x,k_y,k_z), \nonumber\\
\{I \| M_z|\bm{\tau}_{(0,0,1/2)}\}&:\quad s_{x,y,z}(k_x,k_y,k_z) = s_{x,y,z}(k_x,k_y,-k_z), \nonumber\\
\{T U_x(\pi) \| I|\bm 0\}&:\quad s_{y,z}(\bm{k}) = s_{y,z}(-\bm{k}),\; s_{x}(\bm{k}) = -s_{x}(-\bm{k}). \label{eq:const4}
\end{align}
From these constraints, we find that $(s_x(\bm{k}), s_z(\bm{k}))$ forms a reducible two‑dimensional representation $B_{1u} \oplus B_{2g}$ of $D_{2h}$, with basis functions $x$ and $xz$. Expanding each component in a Taylor series around the $\Gamma$ point according to Eq.~(\ref{eq:taylor}) and applying the constraints in Eq.~\ref{eq:const4}, the lowest‑order non‑vanishing terms are
\beqn
s_x(\bm{k}) \propto k_x,\qquad s_z(\bm{k}) \propto k_x k_z.
\eeqn
These leading terms correspond to the basis functions $x$ (of representation $B_{1u}$) and $xz$ (of $B_{2g}$), respectively. Thus, the basis functions of the representation formed by $\bm{S}(\bm{k})$ under $\widetilde{G}$ directly determine the functional form of $\bm{S}(\bm{k})$, as they fully capture its symmetry properties.

\section{B. The parity of the spin textures}
In this section, we elucidate the parity properties of spin textures. As demonstrated in Sec.~A, the non-zero components of $\bm{S}(\bm{k})$ form a representation of the point group $\widetilde{G} = \{\lambda_i R_i\}$. The group $\widetilde{G}$ contains an inversion symmetry $\tilde{P}$ if the SSG $G$ includes any of the symmetries $g_3 = \{ T U_{3} \| I | \bm{\tau}_3 \}$, $g_4 = \{ U_{4} \| P | \bm{\tau}_4 \}$, $g_5 = \{ T \| I | \bm{\tau}_5 \}$, or $g_6 = \{ I \| P | \bm{\tau}_6 \}$; all these operations satisfy $g_{3,4,5,6}\bm{k} = -\bm{k}$. Otherwise, $\widetilde{G}$ lacks inversion.

When $\widetilde{P}$ is realized as $g_5$, it imposes $s_{\alpha}(\bm{k}) = -s_{\alpha}(-\bm{k})$ for all components, enforcing pure odd-parity. When $\widetilde{P}$ is realized as $g_6$, it imposes $s_{\alpha}(\bm{k}) = s_{\alpha}(-\bm{k})$, enforcing pure even-parity. When $\widetilde{P}$ corresponds to $g_3$ or $g_4$, the constraints on $\bm{S}(\bm{k})$ depend on the specific form of the spin rotation $U_i$. Note that if $g_j$ ($j=3,4$) is preserved, then $g_j^2$ is also preserved. Explicitly,
\beqn
g_3^2 = -\{U_3^2 \| I | \bm{\tau}_3^2\},\qquad 
g_4^2 = \{U_4^2 \| I | \bm{\tau}_4^2\},
\eeqn
which lead to
\begin{align}
g_3^2 &: \bm{S}(\bm{k}) = -U_3^2 \bm{S}(\bm{k}), \nonumber\\
g_4^2 &: \bm{S}(\bm{k}) = U_4^2 \bm{S}(\bm{k}).
\end{align}
For a non-zero spin texture ($\bm{S}(\bm{k})\neq \bm{0}$), consistency requires that $U_3$ be a $\pi$ rotation about some axis (so that $U_3^2 = -1$), while $U_4$ may be an arbitrary rotation about some axis. Without loss of generality, we take $U_3 = U_z(\pi)$ and $U_4 = U_z(\theta)$. The symmetries $g_3$ and $g_4$ then impose
\begin{align}
g_3 &: \quad s_{x,y}(\bm{k}) = s_{x,y}(-\bm{k}),\quad s_z(\bm{k}) = -s_z(-\bm{k}), \nonumber\\
g_4(\theta = \pi) &: \quad s_{x,y}(\bm{k}) = -s_{x,y}(-\bm{k}),\quad s_z(\bm{k}) = s_z(-\bm{k}), \nonumber\\
g_4(\theta \neq \pi) &: \quad s_{x,y}(\bm{k}) = 0,\quad s_z(\bm{k}) = s_z(-\bm{k}).
\end{align}
Thus, when $\widetilde{G}$ is a Laue group (i.e., it contains the effective inversion $\widetilde{P}$), each non-zero component of $\bm{S}(\bm{k})$ acquires a definite parity: $s_{\alpha}(\bm{k}) = p_{\alpha} s_{\alpha}(-\bm{k})$ with $p_{\alpha} = \pm 1$. Consequently, for Laue groups, the classification of $\bm{S}(\bm{k})$ into even-parity, odd-parity, and hybrid-parity textures is complete. When $\widetilde{G}$ is not a Laue group, the parity of the spin texture is ill-defined, giving rise to the unconstrained-parity magnets.

\begin{table*}
\centering
\setlength\tabcolsep{9pt}
\renewcommand{\arraystretch}{1.8}
\caption{Classification of even‑parity, odd‑parity, and hybrid‑parity magnets, with symmetry criteria for each type. A check mark (\(\checkmark\)) indicates that the symmetry operation must be present; a cross (\(\times\)) indicates its absence. }
\begin{tabular}{|c|c|c|c|c|c|c|c|c|c|c|}
\hline
Class & Type & Spin texture & \(g_1\) & \(g_2\) & \(g_3\) & \(g_3^{\prime}\) & \(g_4\) & \(g_4^{\prime}\) & \(g_5\) & \(g_6\) \\ \hline
\multirow{5}{*}{Odd‑parity} & \multirow{3}{*}{I} & \multirow{3}{*}{\(s_z(\bm{k}) = -s_z(-\bm{k})\), \(s_{x,y}=0\)} & \(\checkmark\) & \(\times\) & \(\checkmark\) & \(\times\) & \(\times\) & \(\times\) & \(\times\) & \(\times\) \\
\cline{4-11}
 & & & \(\checkmark\) & \(\times\) & \(\times\) & \(\times\) & \(\checkmark\) & \(\times\) & \(\times\) & \(\times\) \\
\cline{4-11}
 & & & \(\checkmark\) & \(\times\) & \(\times\) & \(\times\) & \(\times\) & \(\times\) & \(\checkmark\) & \(\times\) \\
\cline{2-11}
 & II & \(s_{x,y}(\bm{k}) = -s_{x,y}(-\bm{k})\), \(s_z=0\) & \(\times\) & \(\checkmark\) & \(\times\) & \(\times\) & \(\times\) & \(\times\) & \(\checkmark\) & \(\times\) \\
\cline{2-11}
 & III & \(s_{x,y,z}(\bm{k}) = -s_{x,y,z}(-\bm{k})\) & \(\times\) & \(\times\) & \(\times\) & \(\times\) & \(\times\) & \(\times\) & \(\checkmark\) & \(\times\) \\ \hline
\multirow{5}{*}{Even‑parity} & \multirow{3}{*}{I} & \multirow{3}{*}{\(s_z(\bm{k}) = s_z(-\bm{k})\), \(s_{x,y}=0\)} & \(\checkmark\) & \(\times\) & \(\times\) & \(\checkmark\) & \(\times\) & \(\times\) & \(\times\) & \(\times\) \\
\cline{4-11}
 & & & \(\checkmark\) & \(\times\) & \(\times\) & \(\times\) & \(\times\) & \(\checkmark\) & \(\times\) & \(\times\) \\
\cline{4-11}
 & & & \(\checkmark\) & \(\times\) & \(\times\) & \(\times\) & \(\times\) & \(\times\) & \(\times\) & \(\checkmark\) \\
\cline{2-11}
 & II & \(s_{x,y}(\bm{k}) = s_{x,y}(-\bm{k})\), \(s_z=0\) & \(\times\) & \(\times\) & \(\checkmark\) & \(\times\) & \(\times\) & \(\times\) & \(\times\) & \(\checkmark\) \\
\cline{2-11}
 & III & \(s_{x,y,z}(\bm{k}) = s_{x,y,z}(-\bm{k})\) & \(\times\) & \(\times\) & \(\times\) & \(\times\) & \(\times\) & \(\times\) & \(\times\) & \(\checkmark\) \\ \hline
\multirow{3}{*}{Hybrid‑parity} & I & \(s_z(\bm{k}) = -s_z(-\bm{k})\), \(s_x(\bm{k})=s_x(-\bm{k})\), \(s_y=0\) & \(\times\) & \(\times\) & \(\checkmark\) & \(\times\) & \(\checkmark\) & \(\times\) & \(\times\) & \(\times\) \\
\cline{2-11}
 & II & \(s_z(\bm{k}) = -s_z(-\bm{k})\), \(s_{x,y}(\bm{k})=s_{x,y}(-\bm{k})\) & \(\times\) & \(\times\) & \(\checkmark\) & \(\times\) & \(\times\) & \(\times\) & \(\times\) & \(\times\) \\
\cline{2-11}
 & III & \(s_{y,z}(\bm{k}) = -s_{y,z}(-\bm{k})\), \(s_x(\bm{k})=s_x(-\bm{k})\) & \(\times\) & \(\times\) & \(\times\) & \(\times\) & \(\checkmark\) & \(\times\) & \(\times\) & \(\times\) \\ \hline
\end{tabular}
\label{tab:parity_classification}
\end{table*}

\section{C. A unified symmetry classification of even-, odd-, and hybrid-parity magnets}

In this section, we present a unified symmetry classification of even‑parity, odd‑parity, and hybrid‑parity magnets based on the six symmetry operations \(g_1\)–\(g_6\) introduced in the main text. These operations are defined as follows:
\beqn
&&g_1 = \{ U_1\| I | \bm{\tau}_1 \}, \quad 
g_2 = \{ T U_2 \| P | \bm{\tau}_2 \}, \nonumber\\
&&g_3 = \{ TU_3 \| I | \bm{\tau}_3 \}, \quad 
g_4 = \{ U_4 \| P | \bm{\tau}_4 \}, \nonumber\\
&&g_5 = \{ T \| I | \bm{\tau}_5 \}, \quad 
g_6 = \{ I \| P | \bm{\tau}_6 \},
\eeqn
where $g_{1,2}\bm k=\bm k$, and $g_{3,4,5,6}\bm k=-\bm k$. For a non‑zero spin texture \(\bm{S}(\bm{k})\), \(U_1\) and \(U_4\) are constrained to be rotations about some axis by an arbitrary angle, while \(U_2\) and \(U_3\) are constrained to be \(\pi\) rotations about some axis. Without loss of generality, we take
\beqn
U_1 = U_z(\theta),\quad U_2 = U_3 = U_z(\pi),\quad U_4 = U_x(\pi).
\eeqn
For our purpose, we further define $g_3^{\prime}=\{T U_x(\pi)\|I\}$ and $g_4^{\prime}=\{U_z(\pi)\|P|\bm )\}$ with different spin rotation axes from $g_3$
and $g_4$. The constraints of these operations and their primed variants on the spin texture \(\bm{S}(\bm{k})\) are:
\begin{align}
g_1=\{U_z(\theta)\|I|\bm{\tau}_1\}&:\quad \bm{S}(\bm{k}) = (0,0,s_z(\bm{k})), \nonumber\\
g_2=\{T U_z(\pi)\|P|\bm{\tau}_2\}&:\quad \bm{S}(\bm{k}) = (s_x(\bm{k}),s_y(\bm{k}),0),  \nonumber\\
g_3=\{T U_z(\pi)\|I|\bm{\tau}_3\}&:\quad \bm{S}(\bm{k}) = (s_x(-\bm{k}),s_y(-\bm{k}),-s_z(-\bm{k})),  \nonumber\\
g_3^{\prime}=\{T U_x(\pi)\|I|\bm{\tau}_3\}&:\quad \bm{S}(\bm{k}) = (-s_x(-\bm{k}),s_y(-\bm{k}),s_z(-\bm{k})),  \nonumber\\
g_4=\{U_x(\pi)\|P|\bm{\tau}_4\}&:\quad \bm{S}(\bm{k}) = (s_x(-\bm{k}),-s_y(-\bm{k}),-s_z(-\bm{k})),  \nonumber\\
g_4^{\prime}=\{U_z(\pi)\|P|\bm{\tau}_4\}&:\quad \bm{S}(\bm{k}) = (-s_x(-\bm{k}),-s_y(-\bm{k}),s_z(-\bm{k})),  \nonumber\\
g_5=\{T\|I|\bm{\tau}_5\}&:\quad \bm{S}(\bm{k}) = (-s_x(-\bm{k}),-s_y(-\bm{k}),-s_z(-\bm{k})),  \nonumber\\
g_6=\{I\|P|\bm{\tau}_6\}&:\quad \bm{S}(\bm{k}) = (s_x(-\bm{k}),s_y(-\bm{k}),s_z(-\bm{k})).
\label{allcon}
\end{align}
Here \(g_1\) enforces a collinear spin texture (along \(z\)) and \(g_2\) enforces a coplanar texture (in the \(xy\)-plane).  
Operations \(g_3\) and \(g_3^{\prime}\) make one spin component odd (specifically \(s_z\) for \(g_3\), \(s_x\) for \(g_3^{\prime}\)) while the other two remain even. Conversely, \(g_4\) and \(g_4^{\prime}\) make one spin component even (\(s_x\) for \(g_4\), \(s_z\) for \(g_4^{\prime}\)) and the other two odd. These naturally give rise to hybrid‑parity textures when applied alone. However, when combined with a collinear constraint (e.g., \(g_1\)) or a coplanar constraint (e.g., \(g_2\)), the same operations can instead enforce pure even‑ or odd‑parity behavior, because the restricted spin components force all non‑zero components to share the same parity.  The operation \(g_5\) flips both momentum and all spin components, thereby imposing pure odd-parity. In contrast, \(g_6\) flips momentum but leaves all spin components unchanged, forcing pure even-parity.

By combining all the constraints in Eq.~\ref{allcon}, we can classify odd-parity and even-parity magnets each into three types according to the dimensionality of their spin textures: type-I (collinear), type-II (coplanar), and type-III (noncoplanar) \cite{2026Luoxun}. For each parity class, the symmetry conditions that realize these types are summarized in Table~\ref{tab:parity_classification}. 
Specifically, for odd-parity type-I (collinear, \(s_z\) odd), there are three distinct symmetry mechanisms, corresponding to the three rows under odd-parity type-I in the table: they involve \(\{g_1, g_3\}\), \(\{g_1, g_4\}\), and \(\{g_1, g_5\}\), respectively. For odd-parity type-II (coplanar, \(s_{x,y}\) odd), a single mechanism exists, given by the combination \(\{g_2, g_5\}\). Odd-parity type-III (noncoplanar, all components odd) is achieved solely by \(g_5\). Similarly, for even-parity type-I (collinear, \(s_z\) even), three mechanisms appear: \(\{g_1, g_3'\}\), \(\{g_1, g_4'\}\), and \(\{g_1, g_6\}\). Even-parity type-II (coplanar, \(s_{x,y}\) even) is realized by \(\{g_3, g_6\}\), and even-parity type-III (noncoplanar, all components even) by \(g_6\) alone. Hybrid-parity magnets are also classified into three types, each uniquely determined by a single symmetry mechanism: Type-I (coplanar hybrid-parity) with \(s_z\) odd, \(s_x\) even, \(s_y=0\), realized by the combination \(\{g_3, g_4\}\); Type-II (noncoplanar with two even components) with \(s_z\) odd, \(s_{x,y}\) even, realized by \(g_3\) alone; and Type-III (noncoplanar with two odd components) with \(s_{y,z}\) odd, \(s_x\) even, realized by \(g_4\) alone.

Thus, the symmetry framework based on $g_1$--$g_6$ provides a complete and systematic classification of unconventional magnetism across all parity types and spin-texture dimensionalities. Moreover, the established symmetry mechanisms can be realized in different magnetic orders, enabling a wealth of variations of these unconventional magnets \cite{2026Luoxun}. For example, the preservation of symmetries $g_1$ and $g_3^{\prime}$ can be realized both in collinear and in coplanar magnetic orders. Collinear magnetic orders correspond to the widely studied altermagnets. In coplanar magnetic orders, a $d$-wave collinear spin splitting can also be realized in the candidate material CoCrO$_4$ \cite{2025song}.

Notably, the above symmetry considerations of $g_1-g_6$ guarantee only the existence of NSS with a given parity pattern; they do not automatically enforce symmetry-compensated magnetization, which is a defining characteristic of unconventional magnets. For odd‑parity magnets, the condition $s_\alpha(\bm{k}) = -s_\alpha(-\bm{k})$ ensures that states at $\bm{k}$ and $-\bm{k}$ have opposite spin polarizations, so the net magnetization automatically vanishes. This can be verified by checking that the spin point group $\mathcal{P}_s$ (the set of all spin rotations appearing in the SSG) is nonpolar. For example, a collinear odd‑parity magnet generated by $\{g_1, g_3\}$ indeed yields a nonpolar $\mathcal{P}_s$. In contrast, for even‑parity and hybrid‑parity magnets, the parity condition alone ($s_\alpha(\bm{k}) = +s_\alpha(-\bm{k})$ or mixed) does not guarantee zero net magnetization. Additional symmetries with $g\bm{k} \neq \pm \bm{k}$ are required to enforce compensation. Therefore, only when the symmetry criteria for the spin texture are combined with the requirement of a nonpolar $\mathcal{P}_s$ can even‑parity and hybrid‑parity magnets be realized. From this perspective, even‑parity and hybrid‑parity magnets are subject to stricter symmetry constraints than the odd-parity magnets.

\begin{table*}
\centering
\setlength\tabcolsep{6pt}
\renewcommand{\arraystretch}{2}
\caption{Candidate materials for hybrid-parity magnets and unconstrained-parity magnets. Columns 4 and 5 list the dimensionality (collinear, coplanar, noncoplanar, or nonplanar) of the spin textures and magnetic orders in real space, respectively.}
\begin{tabular}{|c|c|c|c|c|}
\hline
Classification & Types & Candidate materials & spin textures & magnetic orders\\
\hline
\multirow{5}{*}{\makecell{\makecell{Hybrid-\\ \\ parity\\ \\ magnets}}}&Type-I& \makecell{FePO$_4$(0.17), NbMnP(0.803),  Fe$_2$WO$_6$(0.812),\\ CsCoBr$_3$(1.0.3),  CsNiCl$_3$
(1.0.4), 
EuIn$_2$As$_2$(1.0.31),\\  RbNiCl$_3$(1.0.34),  CsMnI$_3$(1.0.36), CsMnI$_3$(1.0.37), \\ Ba$_3$MnSb$_2$O$_9$(1.0.46), LaCa$_2$Fe$_3$O$_9$(1.0.30) }&coplanar&coplanar\\
\cline{2-5} 
~&Type-II & \makecell{$\mathrm{ScMnO}_3(0.7)$, $\mathrm{ScMnO}_3(0.8)$,  Ca$_2$CoSi$_2$O$_7$($0.175$), \\ Na$_2$CoP$_2$O$_7$(0.425),  YMnO$_3$(0.6), DyFeO$_3$(0.10), \\ YbMnO$_3$(0.30), HoMnO$_3$(0.31),  YMnO$_3$(0.44), \\Ho$_2$Ge$_2$O$_7$(0.107),  LuFeO$_3$(0.117),  Cu$_3$Mo$_2$O$_9$(0.129),\\ TbFeO$_3$(0.353),    Pb$_2$MnO$_4$(0.552),  HoCrWO$_6$(0.716), \\ TmAgGe(3.1), BaCoSiO$_4$(1.0.49)}&noncoplanar&coplanar \\
\cline{2-5} 
~&Type-III &  FeSb$_2$O$_4$(0.97)&noncoplanar&noncoplanar \\
\hline
\multirow{5}{*}{\makecell{\makecell{Unconstrained \\ \\ -parity\\ \\ magnets}}}&\multicolumn{2}{|c|}{CrSe(2.35),TbCrO$_3$(2.62)} &collinear&noncoplanar \\
\cline{2-5} 
~&\multicolumn{2}{|c|} {\makecell{HoMnO$_3$(0.33), HoMnO$_3$(0.42), HoMnO$_3$(0.43),\\  FeSb$_2$O$_4$(0.97),  SrCo(VO$_4$)(OH)(0.287), \\ DyCrWO$_6$(0.316), Tb$_3$Ge$_5$(0.342),Er$_2$Ge$_2$O$_7$(0.419), \\ 
SrCuTe$_2$O$_6$(0.440),  YbMnO$_3$(0.488), YbMnO$_3$(0.489),\\ SrCuTe$_2$O$_6$(0.530), HoMnO$_3$(0.652),  BaCuTe$_2$O$_6$(0.658), \\ HoCrWO$_6$(0.715), Mn$_3$IrSi(0.898),  Tb$_{14}$Ag$_{51}$(1.0.52),\\ Mn$_3$CoGe(0.900)}}&noplanar&noncoplanar\\
\hline
\end{tabular}
\label{tab3}
\end{table*}

\section{D. Candidate materials for hybrid- and unconstrained-parity magnets} 
In the main text, we have established the symmetry criteria for hybrid-parity magnets and unconstrained-parity magnets. In this section, we present the candidate materials for these classes. In Ref.~\cite{CheXiaobing2024}, the authors used the symmetries that leave the momentum $\bm{k}$ invariant to identify materials exhibiting non-relativistic spin splitting. We take these materials as our input and identify the target symmetry for hybrid-parity spin splitting using the online program FINDSPINGROUP~\cite{CheXiaobing2024}. Moreover, we require that the point group $\mathcal{P}_s$ formed by all spin operations in a SSG is nonpolar, ensuring spin-compensated magnetization. Under these conditions, we find a total of 29 materials for hybrid-parity magnets: 11 Type-I (coplanar spin textures), 17 Type-II (noncoplanar spin textures with two even components), and 1 Type-III (noncoplanar spin textures with two odd components). 

If a material exhibits symmetry-compensated magnetization and non-relativistic spin splitting, but no symmetry relates opposite momenta, we identify it as an unconstrained-parity magnet. Under this criterion, we identify 20 unconstrained-parity magnets: 2 with collinear spin textures and 18 with noncoplanar spin textures. All these materials and their magnetic orders are listed in Table.~\ref{tab3}. We note that the identified type-II HPM include some well-known multiferroic materials, such as YMnO$_3$ and LuMnO$_3$ \cite{Dong02112015}. The coexistence of ferroelectricity and hybrid-parity spin textures in these compounds suggests the possibility of electrically manipulating the spin splitting \cite{DuanXunkai2025,GuMingqiang2025}, paving the way for electrically controlled spintronics devices \cite{2026arXiv260101499Y}.

\section{E. Spin Transport Response Formalism}
	In this section, we present the spin transport response formalism for the spin current and Edelstein effects, covering both symmetry constraints on the response tensors and their numerical evaluation.
	
	\subsection{E1. Symmetry constraints}
	The linear response of the spin current to an applied electric field is described by the third-rank conductivity tensor $\sigma_{ij}^{\alpha}$, defined by $J_i^{\alpha} = \sigma_{ij}^{\alpha} E_j$, where $\alpha = x, y, z$ labels the spin component, $i$ denotes the direction of the spin current, and $j$ the direction of the electric field. 
	Similarly, the Edelstein effect, which describes an electric‑field‑induced spin magnetization $\delta M_i$, is governed by the second-rank susceptibility tensor $\chi_{\alpha j}$ via $\delta M_{\alpha} = - \frac{\mu_{\text B} g_s V}{\hbar} \delta s_{\alpha} = \chi_{\alpha j} E_j$, where $\delta s_i$ is the induced spin polarization, $\mu_{\text B}$ is the Bohr magneton, $g_s \approx 2$ is the electron spin g-factor and $V$ is the volume of the unit cell. 
	
	Both response tensors can be decomposed into time‑reversal‑even ($T$-even) and time‑reversal‑odd ($T$-odd) parts according to their behavior under time reversal. For the spin current conductivity, the $T$-odd component $\sigma_{ij}^{\alpha(o)}$ arises from Fermi‑surface (intra‑band) contributions and depends explicitly on scattering processes, whereas the $T$-even component $\sigma_{ij}^{\alpha(e)}$ originates from Fermi‑sea (inter‑band) contribution and is an intrinsic property of the band structure. In contrast, for the Edelstein susceptibility, the $T$-even component $\chi_{\alpha j}^{(e)}$ corresponds to the  extrinsic Fermi-surface contribution, while the $T$-odd component $\chi_{\alpha j}^{(o)}$ stems from the intrinsic Fermi‑sea mechanism.
	
	Under a spin space group (SSG) operation $g = \{\mathcal{U} \| R|\bm{\tau}\}$, where $\mathcal{U}$ is the spin operation, $R$ is the spatial rotation matrix, and $\bm{\tau}$ a translation vector, these response tensors transform according to the following rules \cite{XIAO2026109872,Etxebarria:cam5007}:
	\beqn
	&&\sigma_{ij}^{\alpha(o)} = \sum_{\beta,m,n} \mathcal{U}_{\alpha\beta} R_{im} R_{jn} \sigma_{mn}^{\beta(o)}, \nonumber\\
	&&\sigma_{ij}^{\alpha(e)} = \det(\mathcal{U}) \sum_{\beta,m,n} \mathcal{U}_{\alpha\beta} R_{im} R_{jn} \sigma_{mn}^{\beta(e)}, \nonumber\\
	&&\chi_{\alpha j}^{(o)} = \sum_{m,n} \mathcal{U}_{im} R_{jn} \chi_{mn}^{(o)}, \nonumber\\
	&&\chi_{\alpha j}^{(e)} = \det(\mathcal{U}) \sum_{m,n} \mathcal{U}_{im} R_{jn} \chi_{mn}^{(e)}. \label{eq:transformation}
	\eeqn
	These transformation laws impose rigorous symmetry constraints that determine which components of $\sigma_{ij}^{\alpha}$ and $\chi_{\alpha j}$ can be non‑zero for a given SSG symmetry.
	
	\subsection{E2. Calculation formalism}
	For numerical evaluation of the response tensors, we employ the standard Kubo formalism of linear response theory. The total spin current conductivity $\sigma_{ij}^{\alpha}$ consists of an intrinsic (Fermi‑sea) part and an extrinsic (Fermi‑surface) part. The intrinsic contribution, which originates from inter‑band transitions and is independent of scattering, is given by \cite{Hu2025}
	\beqn\label{eq:sigma_even}
	\sigma_{ij}^{\alpha(e)} = -\frac{2\hbar e}{V N} \sum_{\bm{k}} \sum_{n \neq m} \frac{\operatorname{Im}\left[ \langle n\bm{k}|\hat{J}_i^{\alpha}|m\bm{k}\rangle \langle m\bm{k}|\hat{v}_j|n\bm{k}\rangle \right]}{(\epsilon_{n\bm{k}} - \epsilon_{m\bm{k}})^2},
	\eeqn
	where the sum runs over occupied bands ($n$) and unoccupied bands ($m$), $e$ is the elementary charge, $\hat{J}_i^{\alpha} = \frac{1}{2}\{\hat{s}_\alpha, \hat{v}_i\}$ is the spin current operator, $\hat{v}_j$ is the velocity operator, $\epsilon_{n\bm{k}}$ the band energies, $V$ the unit cell volume, and $N$ the total number of $\bm{k}$‑points in the Brillouin zone. The extrinsic component, which depends explicitly on the scattering time $\tau$ (or, equivalently, on the broadening parameter $\Gamma = \hbar/\tau$), reads
	\beqn \label{eq:sigma_odd}
	\sigma_{ij}^{\alpha(o)} = -\frac{e\hbar}{\pi V N} \sum_{\bm{k}} \sum_{n,m} \frac{\Gamma^2 \operatorname{Re}\left[ \langle n\bm{k}|\hat{J}_i^{\alpha}|m\bm{k}\rangle \langle m\bm{k}|\hat{v}_j|n\bm{k}\rangle \right]}{\bigl[(E_F-\epsilon_{n\bm{k}})^2+\Gamma^2\bigr]\bigl[(E_F-\epsilon_{m\bm{k}})^2+\Gamma^2\bigr]}.
	\eeqn
	The total spin current conductivity is then $\sigma_{ij}^{\alpha} = \sigma_{ij}^{\alpha(e)} + \sigma_{ij}^{\alpha(o)}$. For the Edelstein effect, the susceptibility tensor $\chi_{\alpha j}$ is obtained by replacing the spin current operator $\hat{J}_i^{\alpha}$ with the spin operator $\hat{s}_i$ in the above expressions. This yields the $T$-even component $\chi_{\alpha j}^{(e)}$ and the $T$-odd component $\chi_{\alpha j}^{(o)}$
	\beqn
	&&\chi_{\alpha j}^{(e)} = \frac{e\mu_{\text B} g_s}{\pi N} \sum_{\bm{k}} \sum_{n,m} \frac{\Gamma^2 \operatorname{Re}\left[ \langle n\bm{k}|\hat{s}_i|m\bm{k}\rangle \langle m\bm{k}|\hat{v}_j|n\bm{k}\rangle \right]}{\bigl[(E_F-\epsilon_{n\bm{k}})^2+\Gamma^2\bigr]\bigl[(E_F-\epsilon_{m\bm{k}})^2+\Gamma^2\bigr]}, \label{eq:chi_even}\\
	&&\chi_{\alpha j}^{(o)} = \frac{2 e \mu_{\text B} g_s}{N} \sum_{\bm{k}} \sum_{n \neq m} \frac{\operatorname{Im}\left[ \langle n\bm{k}|\hat{s}_i|m\bm{k}\rangle \langle m\bm{k}|\hat{v}_j|n\bm{k}\rangle \right]}{(\epsilon_{n\bm{k}} - \epsilon_{m\bm{k}})^2}.
	\eeqn

	\subsection{E3. Equivalence to the semiclassical Boltzmann formalism}
	The Fermi-surface (intra-band) contributions to the spin transport response tensors, $\sigma^{\alpha(o)}_{ij}$ and $\chi^{(e)}_{ij}$, can also be obtained from semiclassical Boltzmann theory under the constant relaxation-time approximation. 
	In the limit $\Gamma \rightarrow 0$, the two formalisms are equivalent, as demonstrated in the following derivation.

	We begin by recalling the Lorentzian function, which approximates the Dirac delta function in the limit $\Gamma \rightarrow 0$,
	\begin{align}\label{eq:Lorentzian}
		\lim_{\Gamma \rightarrow 0} \frac{1}{\pi} \frac{\Gamma}{(x - x_0)^2 + \Gamma^2} = \delta(x - x_0). 
	\end{align}
	Substituting Eq.~\eqref{eq:Lorentzian} into Eq.~\eqref{eq:sigma_odd}, we obtain
	\begin{align}
		\sigma_{ij}^{\alpha(o)} 
		&= -\frac{e\hbar}{V N} \sum_{\bm{k}} \sum_{n,m} \operatorname{Re}\left[ \langle n\bm{k}|\hat{J}_i^{\alpha}|m\bm{k}\rangle \langle m\bm{k}|\hat{v}_j|n\bm{k}\rangle \right] \delta(E_F-\epsilon_{n\bm{k}}) \frac{\Gamma}{(E_F-\epsilon_{m\bm{k}})^2+\Gamma^2}.
	\end{align}
	In the expression above, the term being summed contributes mainly comes from $E_F = \epsilon_{m\bm{k}} = \epsilon_{n\bm{k}}$. Therefore, the expression can be further simplified to
	\begin{align}
		\sigma_{ij}^{\alpha(o)} 
		&= -\frac{e\hbar}{\Gamma V N} \sum_{\bm{k}} \sum_{n} \langle n\bm{k}|\hat{J}_i^{\alpha}|n\bm{k}\rangle \langle n\bm{k}|\hat{v}_j|n\bm{k}\rangle \delta(E_F-\epsilon_{n\bm{k}}) \notag \\
		&= e\tau \sum_{n} \int \frac{d \boldsymbol{k}}{(2\pi)^3} J_{n\boldsymbol{k}}^{i,\alpha} v_{n\boldsymbol{k}}^{j} \frac{\partial f_{n\boldsymbol{k}}^0}{\partial \epsilon_{n\boldsymbol{k}}},
	\end{align}
	where $f_{n\boldsymbol{k}}^0$ denotes the equilibrium Fermi-Dirac distribution. We have use the relation $\Gamma = \hbar/\tau$ and $\frac{\partial f_{n\boldsymbol{k}}^0}{\partial \epsilon_{n\boldsymbol{k}}} = -\delta(E_F-\epsilon_{n\bm{k}})$.

	Through an analogous derivation, Eq.~(\ref{eq:chi_even}) can also be transformed as
	\begin{align}\label{eq:chi_even2}
		\chi_{\alpha j}^{(e)} 
		&= - \frac{e \tau \mu_{\text B}g_s V}{\hbar} \sum_n \int \frac{d \boldsymbol{k}}{(2\pi)^3} s^i_{n\boldsymbol k}v^j_{n\boldsymbol k}  \frac{\partial f_{n\boldsymbol k}^0}{\partial \varepsilon_{n\boldsymbol{k}}},
	\end{align}
	where $s^i_{n\boldsymbol k} = \langle n\bm{k}|\hat{s}_i|n\bm{k}\rangle$ is the spin expectation values.
	
\begin{figure*}[t]
\centering
\includegraphics[width=0.95\textwidth]{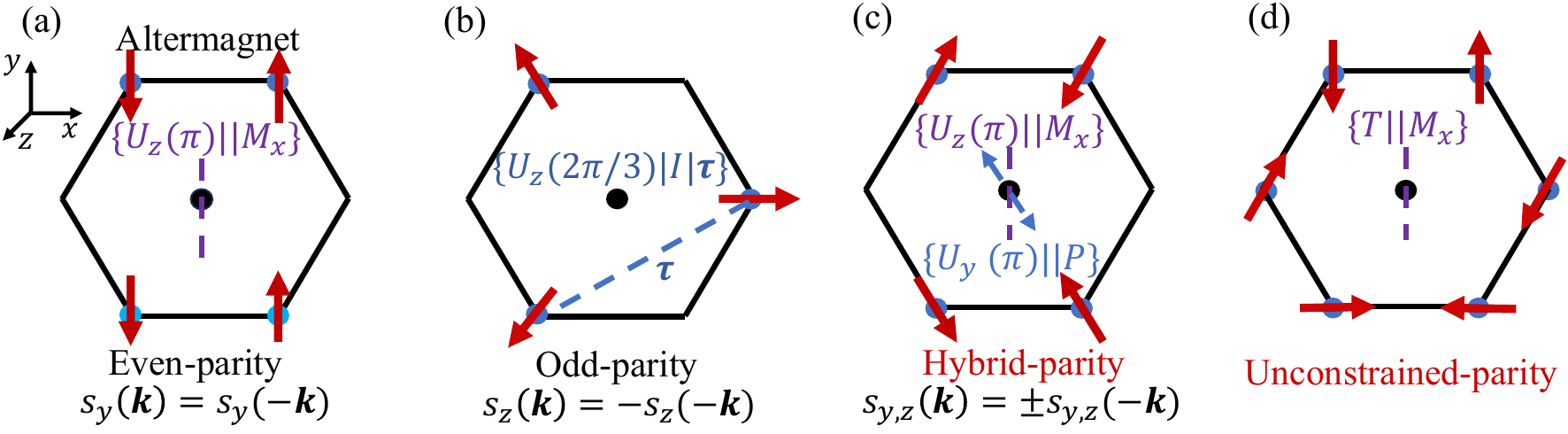}
\caption{Schematic illustration of the four classes of unconventional magnetism on a honeycomb lattice with a $\sqrt{3}\times\sqrt{3}$ supercell. The red arrows denote the magnetic moments, and the governing symmetries are indicated in each panel. (a)~Even-parity altermagnet in a collinear magnetic order, preserving $\{U_z(\pi)\|M_x\}$ and satisfying $s_y(\bm{k})=s_y(-\bm{k})$. (b)~Odd-parity magnet with symmetry $\{U_z(2\pi/3)\|I|\bm{\tau}\}$, characterized by $s_z(\bm{k})=-s_z(-\bm{k})$. (c)~Hybrid-parity magnet preserving $\{U_z(\pi)\|M_x\}$ and $\{U_y(\pi)\|P\}$, where the spin components exhibit mixed parities, $s_{y,z}(\bm{k})=\pm s_{y,z}(-\bm{k})$. (d)~Unconstrained-parity magnet with symmetry $\{T\|M_x\}$, where no symmetry relates $\bm{k}$ and $-\bm{k}$, rendering the spin-texture parity ill-defined.}
\label{Figss}
\end{figure*}

\begin{table}[h!]
\centering
\caption{Summary of magnetic classes, spin texture parities, and non-vanishing response tensor components for the three minimal models. }
\label{tab:responses}
\begin{tabular}{|c|c|c|c|c|}
\hline
Model & Magnetic class & Spin texture parity & $T$-odd spin current & $T$-even Edelstein \\ \hline
$H_1$ & Altermagnet & $s_y(\bm{k}) = s_y(-\bm{k})$ (even) & $\begin{pmatrix}0 & \sigma_{xy}^{y(o)} \\ \sigma_{yx}^{y(o)} & 0\end{pmatrix}$ & None \\ \hline
$H_2$ & Odd-parity magnet & $s_z(\bm{k}) = -s_z(-\bm{k})$ (odd) & None & $\begin{pmatrix}0 & 0 \\ 0 & 0 \\ 0 & \chi_{zy}^{(e)}\end{pmatrix}$ \\ \hline
$H_3$ & Hybrid-parity magnet & $\begin{array}{@{}c@{}} s_y(\bm{k}) = s_y(-\bm{k}) \text{ (even)} \\ s_z(\bm{k}) = -s_z(-\bm{k}) \text{ (odd)} \end{array}$ & $\begin{pmatrix}0 & \sigma_{xy}^{y(o)} \\ \sigma_{yx}^{y(o)} & 0\end{pmatrix}$ & $\begin{pmatrix}0 & 0 \\ 0 & 0 \\ 0 & \chi_{zy}^{(e)}\end{pmatrix}$ \\ \hline
\end{tabular}
\end{table}

\section{F. Theoretical models}

In this section, we construct the theoretical models $H_1$, $H_2$, $H_3$, and $H_4$ to realize the even-parity, odd-parity, hybrid-parity, and unconstrained-parity magnets. These models are defined on a honeycomb lattice with a $\sqrt{3}\times\sqrt{3}$ magnetic supercell, as shown in Fig.~\ref{Figss}. We investigate their spin textures and study the Fermi-surface-mediated spin transport properties in models $H_{1,2,3}$.

\subsection{F1. Model Hamiltonians and spin textures}
The Hamiltonian for all three models adopts a unified tight-binding form
\begin{equation}
H_{1,2,3} = \sum_{\langle ij\rangle,\sigma} t \, c_{i\sigma}^\dagger c_{j\sigma} + J \sum_{i,\sigma,\sigma'} \bm{m}_i \cdot c_{i\sigma}^\dagger \bm{\sigma}_{\sigma\sigma'} c_{i\sigma'} - \sum_{i\in{\alpha},\sigma} \delta_{i} c_{i\sigma}^\dagger c_{i\sigma},
\end{equation}
where $t$ is the nearest-neighbor hopping amplitude, $J$ the exchange coupling strength, and $\bm{m}_i$ a unit vector specifying the orientation of the local magnetic moment on site $i$. Each model encompasses six sublattices labeled $a,b,c,d,e,f$ arranged in counterclockwise order and the last term represents a sublattice-dependent potential, with $\alpha$ labeling the sublattice index.  The specific magnetic configurations and $\mu_{\alpha}$ are:
\begin{itemize}
\item \textbf{Model $H_1$ (even-parity altermagnet):}  
  $\bm{m}_a=(0,-1,0)$, $\bm{m}_b=(0,1,0)$, $\bm{m}_d=(0,1,0)$, $\bm{m}_e=(0,-1,0)$, and $\bm{m}_c=\bm{m}_f=\bm{0}$.  For our purpose, we take $\delta_{a}=\delta_{b}=\delta_{c}=\delta_{f}=0$ and $\delta_d=\delta_e=1$ to break the $PT$ symmetry.
In this case, the system features spin-contrasted sublattices related by mirror symmetry along the $x$-direction. The resulting spin texture is collinear along the $y$-axis and satisfies $s_y(\bm{k}) = s_y(-\bm{k})$, characteristic of an even-parity altermagnet.

\item \textbf{Model $H_2$ (odd-parity magnet):}  
  $\bm{m}_a=(-1/2,-\sqrt{3}/2,0)$, $\bm{m}_c=(1,0,0)$, $\bm{m}_e=(1/2,\sqrt{3}/2,0)$, $\bm{m}_b=\bm{m}_d=\bm{m}_f=\bm{0}$, and $\delta_{\alpha}=0$ for all sublattices.  
  This coplanar magnetic configuration preserves the symmetry operations $g=\{U_z(2\pi/3)\|I|\bm{\tau}\}$ (with $\bm{\tau}=(0,\sqrt{3})a$, where $a$ is the lattice constant) and $g_3=\{TU_z(\pi)\|I|\bm 0\}$. These symmetries enforce a collinear spin texture along the $z$-direction with odd-parity: $s_z(\bm{k}) = -s_z(-\bm{k})$.

\item \textbf{Model $H_3$ (Type-I hybrid-parity magnet):}  
  $\bm{m}_a=(1/2,-\sqrt{3}/2,0)$, $\bm{m}_b=(-1/2,\sqrt{3}/2,0)$, $\bm{m}_d=-(1/2,\sqrt{3}/2,0)$, $\bm{m}_e=(1/2,\sqrt{3}/2,0)$, $\bm{m}_c=\bm{m}_f=\bm{0}$, and $\delta_{\alpha}=0$ for all sublattices.  
  This coplanar order respects the symmetries $g_4=\{U_y(\pi)\|P|\bm 0\}$ and $g_3=\{TU_z(\pi)\|I|\bm 0\}$, which together generate a hybrid-parity spin texture featuring both even and odd components: $s_y(\bm{k}) = s_y(-\bm{k})$ (even-parity) and $s_z(\bm{k}) = -s_z(-\bm{k})$ (odd-parity), with $s_x(\bm{k}) \equiv 0$.

\item \textbf{Model $H_4$ (unconstrained-parity magnet):}  
  The magnetic moments are arranged noncoplanarly:  
  $\bm{m}_a=(1,0,0)$, $\bm{m}_b=(-1,0,0)$, $\bm{m}_c=(0,1,0)$, $\bm{m}_d=(0,0,1)$, $\bm{m}_e=(0,0,-1)$, $\bm{m}_f=(0,-1,0)$. 
  Site potentials are set as $\delta_a=\delta_b=0$, $\delta_c=\delta_f=t$, and $\delta_d=\delta_e=-t$.  
  This configuration breaks all symmetries that relate $\bm{k}$ and $-\bm{k}$, so the parity of the spin texture is ill‑defined. Nevertheless, the preserved symmetry $\{T\|M_x|\bm{0}\}$ enforces symmetry‑compensated magnetization. Non‑relativistic spin splitting is present, and the magnitude of the spin expectation value can differ at $\bm{k}$ and $-\bm{k}$ (i.e., $|s_\alpha(\bm{k})| \neq |s_\alpha(-\bm{k})|$ for $\alpha=x,y,z$), as shown in Fig.~\ref{Figun}. This behavior is characteristic of unconstrained‑parity magnets.
  
\end{itemize}

The spin textures on the Fermi surface of models $H_1-H_3$ are shown in Fig.~\ref{Figs1}(a)-\ref{Figs1}(d) for all three models. For $H_1$, the spin polarization is collinear along $y$ and exhibits even-parity as expected. Moreover, $s_y(k_x,k_y) = -s_y(-k_x,k_y)$ due to the symmetry $\{U_z(\pi)|M_x|\bm 0\}$, with $M_x$ being spatial mirror. For $H_2$, the texture is collinear along $z$ and satisfies $s_z(R_6\bm{k}) = -s_z(\bm{k})$ due to the symmetries $\{U_z(2\pi/3)\|R_3|\bm 0\}$ and $\{TU_z(\pi)|I|\bm 0\}$ (where $R_6$ and $R_3$ denote sixfold and threefold spatial rotations, respectively), realizing an $f$-wave magnet. Moreover, $s_z(k_x,k_y)=-s_z(k_x,-k_y)$ due to the existence of symmetry $\{U_x(\pi)||M_y|\bm 0\}$.
For $H_3$, the texture possesses two non-vanishing components with distinct parity properties: $s_y(\bm{k}) = s_y(-\bm{k})$ (even) and $s_z(\bm{k}) = -s_z(-\bm{k})$ (odd), while $s_x(\bm{k}) = 0$. Moreover, $s_y(k_x,k_y) = -s_y(-k_x,k_y)$ and $s_z(k_x,k_y) = s_z(-k_x,k_y)$ due to the existence of symmetry $\{U_z(\pi)\|M_x|\bm 0\}$.

\begin{figure*}[t]
\centering
\includegraphics[width=0.95\textwidth]{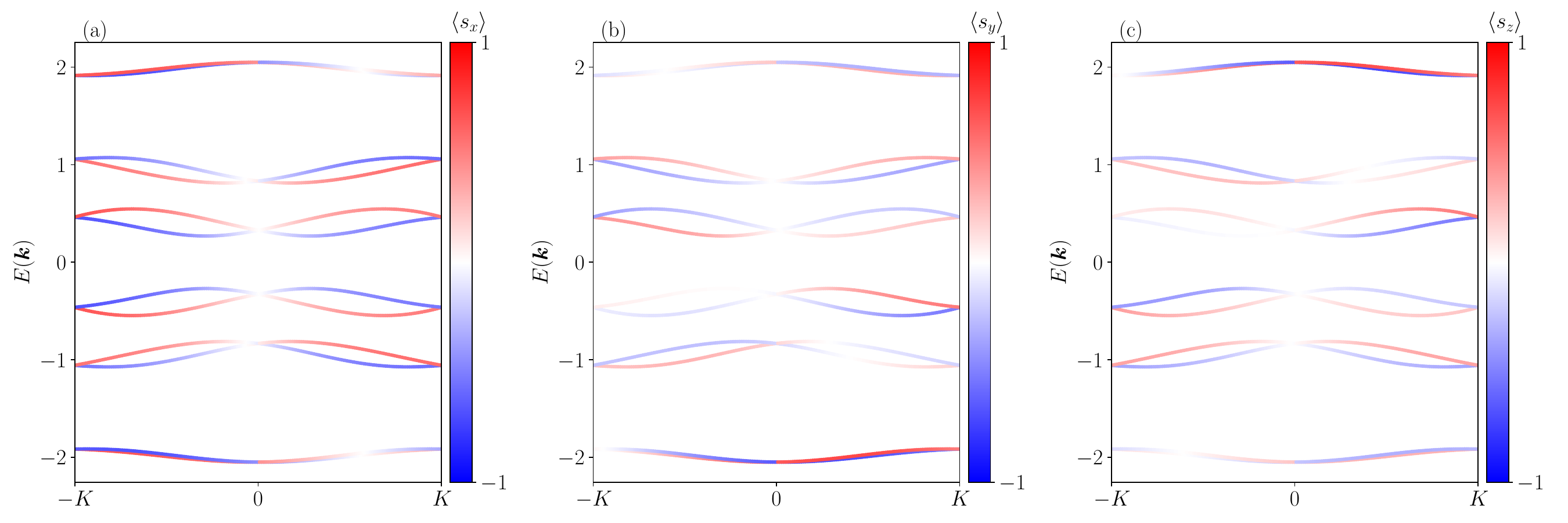}
\caption{Electronic band structure of a prototypical unconstrained-parity magnet along a high-symmetry path from $-K$ to $K$. $K=(2\pi/(3\sqrt{3}a),2\pi/3a)$ with $a$ being the lattice constant. The color scale indicates the spin expectation values $\langle s_\alpha\rangle$ ($\alpha=x,y,z$) for each band in panels (a)-(c).  }
\label{Figun}
\end{figure*}

\subsection{F2. Symmetry constraints on response tensors}

All models $H_{1,2,3}$ are two-dimensional with the lattice confined to the $xy$-plane, restricting the electric field to components $E_x$ and $E_y$ only. For a fixed spin component $\alpha$, the spin current tensor $\sigma_{ij}^{\alpha}$ constitutes a $2\times2$ matrix in spatial indices ($i,j = x,y$). The Edelstein susceptibility tensor $\chi_{\alpha j}$ forms a $3\times2$ matrix with spin indices $\alpha = x,y,z$ and spatial indices $j = x,y$. Applying the transformation rules in Eq.~(\ref{eq:transformation}) together with the preserved symmetries of each model, we derive the allowed non-zero tensor components, presented explicitly below in matrix form.

\begin{itemize}
\item \textbf{Model $H_1$ (altermagnet):}  
$H_1$ respects the symmetries $\{U_z(\pi)\|M_x|\bm 0\}$,  $\{TU_z(\pi)\|I|\bm 0\}$, and continuous spin rotation $\{U_y(\theta)\|I|\bm 0\}$ for arbitrary $\theta$. These symmetries impose the following constraints on the response tensors. For the $T$-odd spin current $\sigma^{\alpha(o)}$, only the $\alpha = y$ component survives, with the form:
 \beqn
\sigma^{y(o)} = \begin{pmatrix} 0 & \sigma_{xy}^{y(o)} \\ \sigma_{yx}^{y(o)} & 0 \end{pmatrix},
\eeqn
  while all other spin components $\sigma^{\alpha(o)} = \bm{0}$ (zero $2\times2$ matrices). The $T$-even spin current vanishes identically: $\sigma^{\alpha(e)} = \bm{0}$ for $\alpha = x,y,z$. Both Edelstein susceptibility components are forbidden: $\chi^{(e)} = \chi^{(o)} = \bm{0}$ (zero $3\times2$ matrices).

\item \textbf{Model $H_2$ (odd-parity magnet):}  $H_1$ respects the symmetries  $\{U_z(2\pi/3)\|I|\bm{\tau}\}$, $\{TU_z(\pi)\|I|\bm 0\}$, and $\{U_x(\pi)\|M_y|\bm 0\}$. These operations yield the following constraints. The $T$-odd spin current vanishes completely: $\sigma^{\alpha(o)} = \bm{0}$ for all $\alpha$. For the $T$-even spin current, only the $\alpha = z$ component is permitted, with the form:
  \[
  \sigma^{z(e)} = \begin{pmatrix} 0 & \sigma_{xy}^{z(e)} \\ \sigma_{yx}^{z(e)} & 0 \end{pmatrix}.
  \]
 For the Edelstein effect, only the $T$-even component (Fermi-sea origin) is allowed to be non-zero with the form:
\beqn
  \chi^{(e)} = \begin{pmatrix} 0 & 0 \\ 0 & 0 \\ 0 & \chi_{zy}^{(e)} \end{pmatrix},
\eeqn
 and the $T$-odd Edelstein susceptibility vanishes: $\chi^{(o)} = \bm{0}$.

\item \textbf{Model $H_3$ (hybrid-parity magnet):} $H_3$ respects the symmetries
 $\{U_z(\pi)\|M_x|\bm 0\}$, $\{TU_z(\pi)\|I|\bm 0\}$, and $\{U   _y(\pi)\|P|\bm 0\}$. These symmetries lead to the following response tensor structure. For the $T$-odd spin current, only the $\alpha = y$ component is allowed, with the form:
\beqn
  \sigma^{y(o)} = \begin{pmatrix} 0 & \sigma_{xy}^{y(o)} \\ \sigma_{yx}^{y(o)} & 0 \end{pmatrix}.
\eeqn
  The $T$-even spin current vanishes: $\sigma^{\alpha(e)} = \bm{0}$ for $\alpha = x,y,z$. Notably, the absence of $T$-even spin current is a general characteristic of type-I HPM. Because the symmetry $\{TU_z(\pi)\|I|\bm 0\}$ enforces $\sigma^{x,y(e)}=0$ and the symmetry $\{U_y(\pi)\|P|\bm 0\}$ enforces $\sigma^{x,z(e)}=0$, while these two symmetry are typically preserved in type-I HPMs. For the Edelstein effect, both $T$-odd and $T$-even components are permitted. The $T$-odd susceptibility (Fermi-sea origin) takes the form:
\beqn
  \chi^{(o)} = \begin{pmatrix} \chi_{xx}^{(o)} & 0 \\ 0 & 0 \\ 0 & 0 \end{pmatrix}.
\eeqn
  The $T$-even Edelstein susceptibility (Fermi-surface origin) takes the form:
\beqn
  \chi^{(e)} = \begin{pmatrix} 0 & 0 \\ 0 & 0 \\ 0 & \chi_{zy}^{(e)} \end{pmatrix}.
\eeqn
\end{itemize}
The $T$-odd spin current and $T$-even Edelstein responses are summarized in Table~\ref{tab:responses}, which additionally lists the spin texture parity properties for each model. The correspondence between spin texture parity and Fermi-surface response properties is consistent: even-parity spin textures enable $T$-odd spin currents (magnetic spin Hall effect), odd-parity textures allow $T$-even Edelstein responses, while hybrid-parity textures enable both types of responses simultaneously.

\begin{figure*}[t]
\centering
\includegraphics[width=0.95\textwidth]{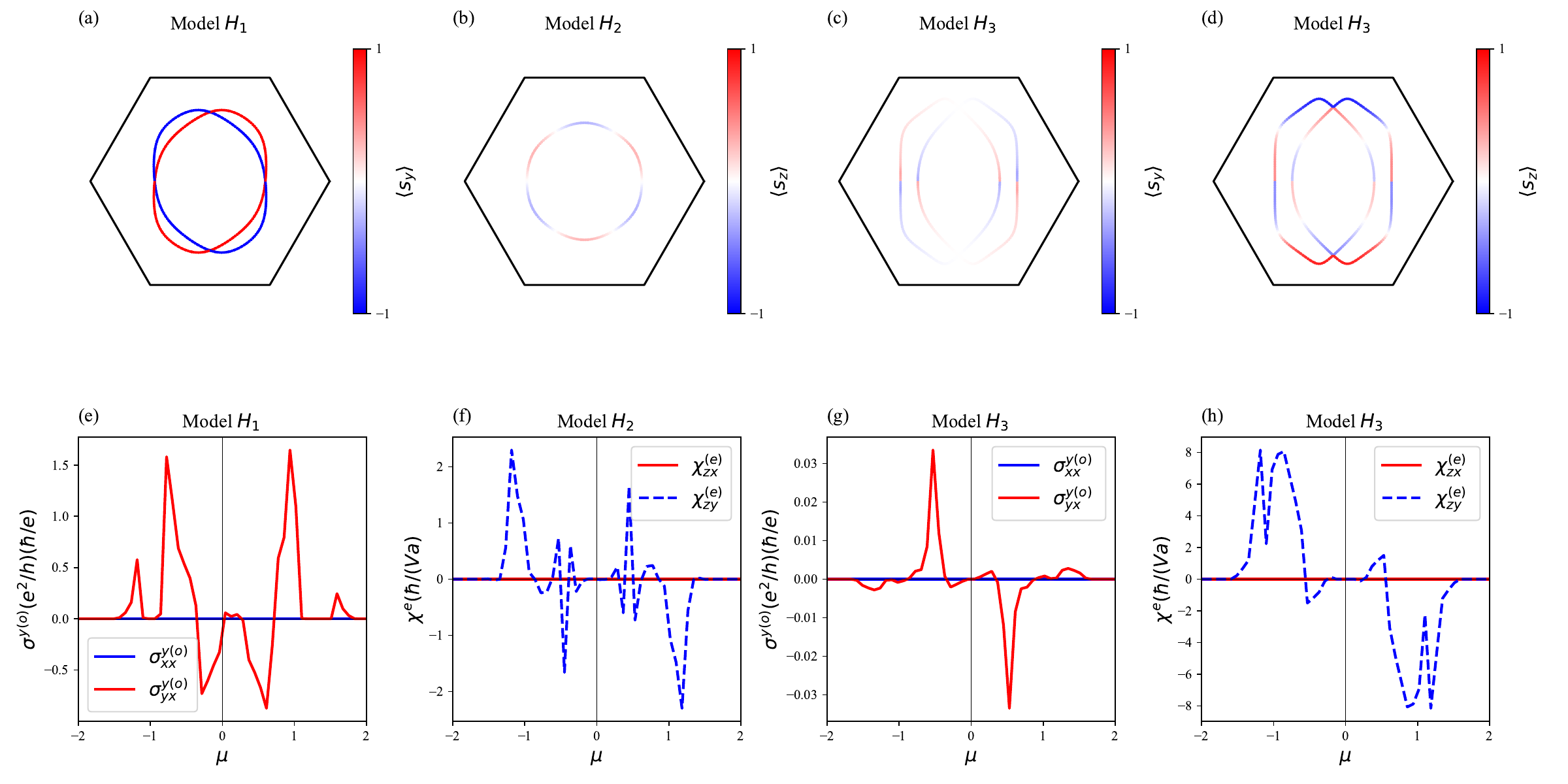}
\caption{Fermi-surface spin textures and transport responses for models $H_1$, $H_2$, and $H_3$. (a)–(d) Spin textures $\bm{S}(\bm{k})$ on the Fermi surface: (a) $H_1$: collinear along $y$ with even-parity $s_y(\bm{k})=s_y(-\bm{k})$; (b) $H_2$: collinear along $z$ with odd-parity $s_z(\bm{k})=-s_z(-\bm{k})$; (c) and (d) $H_3$: $s_y(\bm{k})$ (even) and $s_z(\bm{k})$ (odd), illustrating hybrid parity. (e) Model $H_1$: $T$-odd spin Hall conductivity $\sigma_{xy}^{y(o)}$ (red solid) as a function of $\mu$, while $\sigma_{xx}^{y(o)}$ (blue dashed) vanishes. (f) Model $H_2$: $T$-even Edelstein susceptibility $\chi_{zy}^{(e)}$ (blue dashed); $\chi_{zx}^{(e)}$ (red solid) vanishes. (g) Model $H_3$: $T$-odd spin Hall conductivity $\sigma_{yx}^{y(o)}$ (red solid), arising from the even-parity $s_y$ component. (h) Model $H_3$: $T$-even Edelstein susceptibility $\chi_{zy}^{(e)}$ (blue dashed), arising from the odd-parity $s_z$ component.  All results agree with symmetry-imposed selection rules. In (e)-(h), we take $\Gamma=0.01$eV.}
\label{Figs1}
\end{figure*}

\subsection{F3. $T$-odd spin current and $T$-even Edelstein effect}

We perform tight-binding model calculations to study the Fermi-surface-mediated spin transport properties for models $H_{1,2,3}$, using the formalism outlined in Sec.~E2 with hopping parameter $t=1$ eV and exchange coupling $J=0.5$ eV. The numerical results fully confirm the symmetry-imposed selection rules derived in Sec.~E2.

For model $H_1$, the $T$-odd spin current components $\sigma_{xy}^{y(o)}$ (red solid) and $\sigma_{xx}^{y(o)}$ (blue solid) as functions of chemical potential $\mu$ are plotted in Fig.~\ref{Figs1}(e): $\sigma_{xy}^{y(o)} \neq 0$ while $\sigma_{xx}^{y(o)} = 0$, consistent with the predicted tensor structure by symmetry and realizing a magnetic spin Hall effect. For model $H_2$, the $T$-even Edelstein susceptibilities $\chi_{zx}^{(e)}$ (red solid) and $\chi_{zy}^{(e)}$ (blue dashed) are shown in Fig.~\ref{Figs1}(f): only $\chi_{zy}^{(e)} \neq 0$, while $\chi_{zx}^{(e)} = 0$, consistent with spin textures $s_z(k_x,k_y) = -s_z(k_x,-k_y)$ and $s_z(k_x,k_y) = s_z(-k_x,k_y)$.
For model $H_3$, both $T$-odd spin current and $T$-even Edelstein responses coexist: Fig.~\ref{Figs1}(g) shows the $T$-odd spin current $\sigma_{yx}^{y(o)}$ (red solid); Fig.~\ref{Figs1}(h) shows the Edelstein susceptibilities $\chi_{zy}^{(e)}$ (blue dashed lines).

	\begin{figure}[htbp]
		\begin{centering}
			\includegraphics[width=\textwidth]{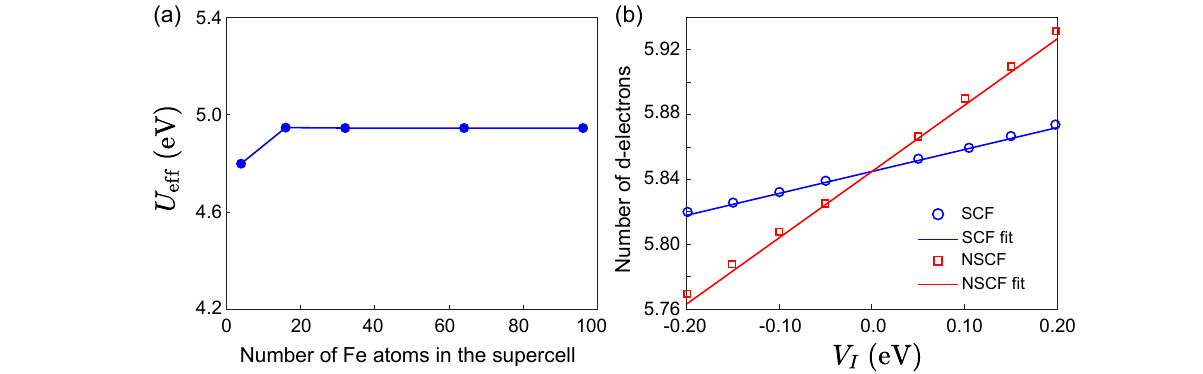}
			\par\end{centering}
		\caption{
			Determination of the effective parameter $U_{\mathrm{eff}}$ for Fe-$3d$ orbitals in \ch{FePO4} using the linear response method. (a) Calculated $U_{\mathrm{eff}}$ as a function of supercell size, represented by the total number of Fe atoms. The calculated value converges to $4.9$~eV in the largest $4\times 2\times 3$ supercell. (b) Linear response of the $d$-electron occupation number $N_I$ to an applied localized potential $V_I$ for the $4\times2\times3$ supercell.
		}
		\label{fig:figS1} 
	\end{figure}

	\begin{figure}[htbp]
		\begin{centering}
			\includegraphics[width=\textwidth]{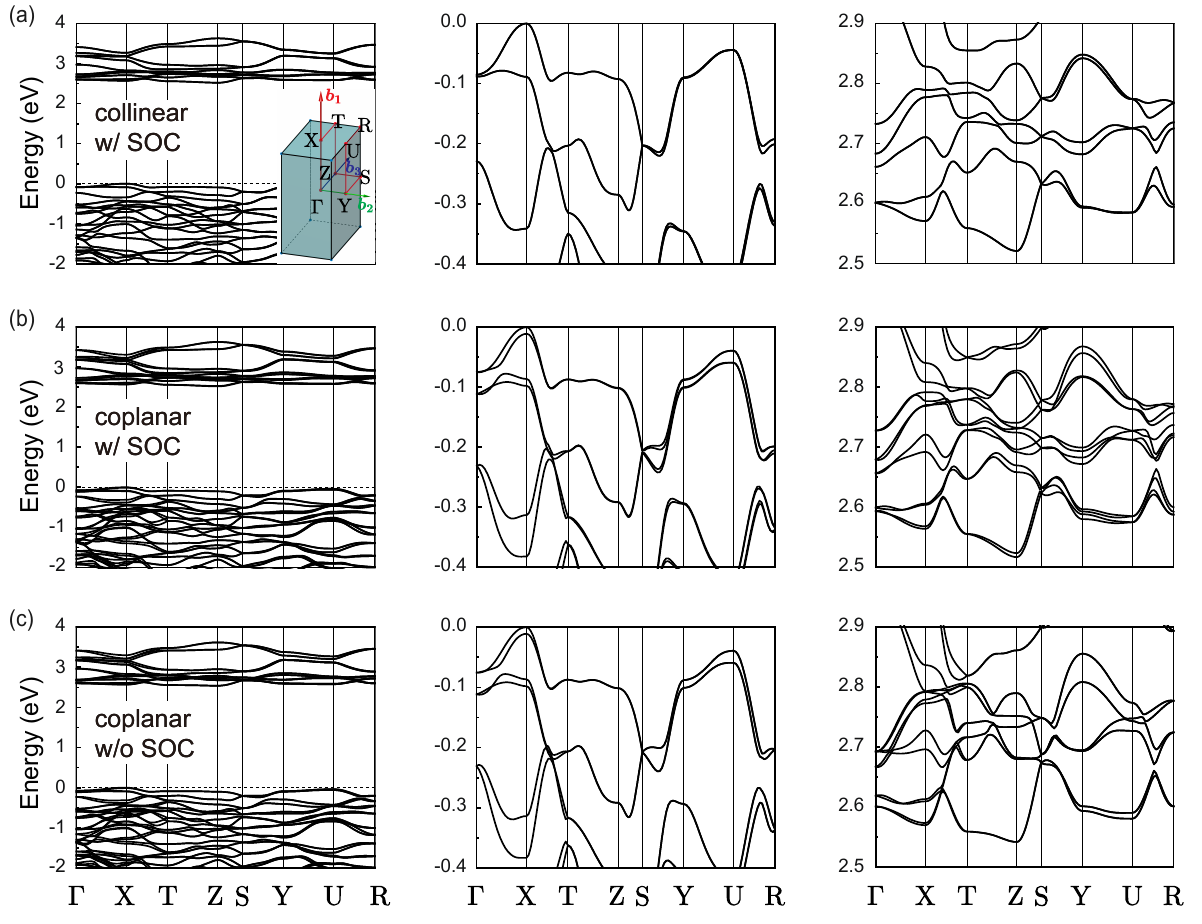}
			\par\end{centering}
		\caption{
		Calculated electronic band structures of \ch{FePO4} under different magnetic configurations. Results are shown for (a) the collinear magnetic order with SOC, (b) the coplanar magnetic order with SOC, and (c) the coplanar magnetic order without SOC. The first column presents the band structures over a broad energy range. The second and third columns display magnified views of the energy bands near the valence band maximum (from $-0.4$ to $0.0$~eV) and the conduction band minimum (from $2.5$ to $2.9$~eV), respectively. The highest occupied state is set to $0$~eV, indicated by the horizontal dashed lines.
		}
		\label{fig:figS2} 
	\end{figure}

\section{G. First-principles calculations}
	\subsection{G1. Computation methods}
	Density functional theory (DFT) calculations involved in this work were implemented in the Vienna ab initio Simulation Package (VASP) code \cite{vasp1,vasp2}. 
	The exchange and correlation interactions were described by Perdew-Burke-Ernzerhof (PBE) \cite{GGA} functional based on the generalized gradient approximation (GGA).
	Projector-augmented wave (PAW) method  \cite{PAW} was adopted and the kinetic energy cutoff was set to be 600 eV. The first Brillouin zone (BZ) was sampled by Monkhorst-Pack meshes method \cite{MPsampling} with a 8$\times$4$\times$7 $k$-point grid for primitive cell. 
	The lattice parameters were fixed at experimentally determined values (\ie $a = 4.7560~\text{\AA}$, $b = 9.7599~\text{\AA}$, $c = 5.7519~\text{\AA}$, \cite{rousse2003magnetic}), and the atomic positions were fully relaxed.
	The force and energy convergence criterion was set to  $0.01~\text{eV/\AA}$ and $10^{-6} \ \rm{eV}$, respectively. 
	The simplified rotationally invariant approach \cite{PhysRevB.57.1505} was used to describe the strongly correlated $d$  electrons of the Fe atoms in \ch{FePO4}. 
	The effective on-site Coulomb interaction $U_{\mathrm{eff}} = 4.9~\mathrm{eV}$ was determined using the linear response approach introduced by M. Cococcioni and S. de Gironcoli \cite{PhysRevB.71.035105}. 
	A coplanar magnetic structure was set for the electronic structure calculations, with the direction of the magnetic moments constrained to align with the experimentally measured results \cite{rousse2003magnetic}.
	
	To calculate the spin transport response tensors, a Wannier tight binding Hamiltonian consisting of Fe-$3d$ and O-$2p$ orbitals was constructed using the Wannier90 package \cite{wannier90_2020}.
	The momentum-space integrals specified in Eqs.~(\ref{eq:sigma_even})-(\ref{eq:chi_even2}) were then evaluated with an in-house parallelized Julia code. 
	The first BZ  was sampled with a $k$-mesh $160\times 80\times140$ and the convergence is well tested with a denser $k$-mesh $200\times 100\times175$. 
	The Dirac delta function was approximated by Gaussian function with a broadening width of 50 meV. 
	The relaxation time $\tau$ was set to $20~\mathrm{fs}$, which is a moderate value compared to common metals \cite{metal_relaxtion_time} and metal oxides \cite{10.1063/1.3562141,PhysRevB.95.205202}.

	\subsection{G2. The linear response calculation for $U_{\mathrm{eff}}$}
	To account for the strong correlation effects arising from the localized Fe-$3d$ electrons in \ch{FePO4}, we employ the linear-response
	method \cite{PhysRevB.71.035105} within the VASP framework to determine the effective parameter $U_{\mathrm{eff}}$. 
	This parameter is given by the self-consistent response coefficient $\chi_{\mathrm{SCF}}$ and the non-self-consistent response coefficient $\chi_{\mathrm{NSCF}}$.
	\begin{align}
		U_{\mathrm{eff}} = \chi_{\text{SCF}}^{-1} - \chi_{\text{NSCF}}^{-1} \approx \left(\frac{\partial N_I^{\mathrm{SCF}}}{\partial V_I}\right)^{-1} - \left(\frac{\partial N_I^{\mathrm{NSCF}}}{\partial V_I}\right)^{-1},
	\end{align}
	where $V_I$ is an additional spherical potential applied to the $d$-orbital manifold of the $I$-th Fe atom. $N_I^{\mathrm{SCF}}$ and $N_I^{\mathrm{NSCF}}$ denote the number of $d$ electrons localized on site $I$, obtained from self-consistent and non-self-consistent calculations, respectively.
	
	To eliminate the artificial coupling between the perturbed atom $I$ and its periodic replicas, we performed a convergence test by constructing supercells of increasing size (\ie 1$\times$1$\times$1, 2$\times$1$\times$2, 2$\times$2$\times$2, 4$\times$2$\times$2, and 4$\times$2$\times$3)\cite{PhysRevB.71.035105}. 
	The results are presented and plotted in Fig.~\ref{fig:figS1}(a), yielding a converged effective parameter of $U_{\mathrm{eff}} = 4.9~\mathrm{eV}$, a value that is consistent with previous literature\cite{zhou2004electronic,calderon2015aflow}.
	Fig.~\ref{fig:figS1}(b) shows the linear response of $N_{I}$ and $V_I$ in a $4\times 2\times 3$ supercell, calculated both self-consistently and non-self-consistently.
	With the inclusion of the effective parameter $U_{\mathrm{eff}}$, our DFT+$U$ calculation yields magnetic moments for the Fe atoms in \ch{FePO4} of $|M_x| = 0.00~\mu_{\mathrm B}$, $|M_y| = 4.24~\mu_{\mathrm B}$, and $|M_z| = 0.94~\mu_{\mathrm B}$,  consistent with the experimental measurements reported in Ref.~\cite{rousse2003magnetic}.

\subsection{G3. Band structures, spin textures, and spin transport}

We first examine the electronic band structure of bulk FePO$_4$. For a collinear magnetic alignment with moments along $z$, the system preserves $PT$ symmetry, leading to spin-degenerate bands [Fig.~\ref{fig:figS2}(a)]. In the actual coplanar magnetic ground state, however, $PT$ symmetry is broken, and non-relativistic spin splitting emerges even without spin-orbit coupling (SOC) [Fig.~\ref{fig:figS2}(b)]. Including SOC [Fig.~\ref{fig:figS2}(c)] causes only minor modifications, confirming that the spin splitting is predominantly of non-relativistic origin. Band structures over a wide energy range, as well as magnified views near the valence band maximum and conduction band minimum, are provided in Fig.~\ref{fig:figS2}.

The hybrid-parity spin texture of FePO$_4$ is visualized in Fig.~\ref{fig:figS3}. Without SOC [Figs.~\ref{fig:figS3}(a) and \ref{fig:figS3}(b)], the spin components in the plane are shown by arrows, while the color map represents the spin component $s_x(\bm{k})$ in (a) and $s_z(\bm{k})$ in (b). The texture clearly displays the even-parity $s_z$ and odd-parity $s_x$ components, consistent with the symmetry analysis. When SOC is turned on [Figs.~\ref{fig:figS3}(c) and \ref{fig:figS3}(d)], the overall structure remains essentially unchanged, indicating that SOC plays a minor role in shaping the Fermi-surface spin texture.

The coplanar antiferromagnetic ground state of FePO$_4$ belongs to the magnetic space group $P2_12_12_1$. The corresponding magnetic point group is $222.1$, generated by the twofold rotations $C_{2x}$, $C_{2y}$, and $C_{2z}$. Under the constraints of this magnetic point group, both the $\mathcal{T}$-even and $\mathcal{T}$-odd magnetoelectric tensors take identical diagonal forms,
\begin{equation}
\alpha^{(e)}=\alpha^{(o)}=
\begin{pmatrix}
\alpha_{xx} & 0 & 0 \\
0 & \alpha_{yy} & 0 \\
0 & 0 & \alpha_{zz}
\end{pmatrix},
\end{equation}
while the $\mathcal{T}$-even and $\mathcal{T}$-odd spin Hall conductivity tensors share identical symmetry-adapted structures, with non-zero elements distributed across all three spin-polarization channels:
\begin{align}
\sigma^{x(e)}=\sigma^{x(o)}&=
\begin{pmatrix}
0 & 0 & 0\\
0 & 0 & \sigma^{x}_{yz}  \\
0 & \sigma^{x}_{zy} & 0
\end{pmatrix},\nonumber\\
\sigma^{y(e)}=\sigma^{y(o)}&=
\begin{pmatrix}
0 & 0 & \sigma^{y}_{xz} \\
0 & 0 & 0 \\
\sigma^{y}_{xz} & 0 & 0
\end{pmatrix}, \nonumber\\
\sigma^{z(e)}=\sigma^{z(o)}&=
\begin{pmatrix}
0 & \sigma^{z}_{xy} & 0 \\
\sigma^{z}_{yx} & 0 & 0 \\
0 & 0 & 0
\end{pmatrix}. \nonumber\\
\end{align}
We compute the Fermi-energy dependence of the spin transport response tensors (Fig.~\ref{fig:figS4}). The left and right columns present the results without and with spin-orbit coupling (SOC), respectively. Panels (a) and (b) display the $\mathcal{T}$-odd and $\mathcal{T}$-even spin Hall conductivities, $\sigma_{ij}^{s(o)}$ and $\sigma_{ij}^{s(e)}$, while panels (c) and (d) show the $\mathcal{T}$-even and $\mathcal{T}$-odd Edelstein susceptibilities, $\chi_{ij}^{(e)}$ and $\chi_{ij}^{(o)}$. Different tensor components are distinguished by colors as indicated in the legends.

\section{H. Response tensors in altermagnets and odd-parity magnets}

In this section, we demonstrate the absence of the $T$-even Edelstein effect in altermagnets and the absence of the $T$-odd spin current in odd-parity magnets.

\subsection{H1. Absence of the Edelstein effect in altermagnets}

Altermagnets are associated with collinear magnetic orders (assumed along the $z$ direction), where spin-contrasted sublattices are related by rotation or mirror operations. They typically preserve the symmetries $\{U_z(\theta)\|I|\bm{0}\}$ and $\{TU_x(\pi)\|I|\bm{0}\}$. The $T$-even Edelstein susceptibility $\chi_{\alpha j}^{(e)}$ transforms under a SSG operation $\{\mathcal{U}\|R|\bm \tau\}$ as
\beqn
\chi_{\alpha j}^{(e)} = \det(\mathcal{U}) \sum_{m,n} \mathcal{U}_{\alpha m} R_{jn} \chi_{mn}^{(e)}.
\eeqn
For symmetry $\{U_z(\theta)\|I|\bm{0}\}$, we have $R=I$ and $\det(U_z)=1$. For a fixed spatial index $j$, the spin vector $(\chi_{xj}^{(e)},\chi_{yj}^{(e)},\chi_{zj}^{(e)})$ must be invariant under arbitrary rotation $U_z(\theta)$. The only invariant direction is the $z$ axis, hence $\chi_{xj}^{(e)}=\chi_{yj}^{(e)}=0$, leaving only $\chi_{zj}^{(e)}$ possibly non-zero. For symmetry $\{TU_x(\pi)\|I|\bm{0}\}$, we have
\beqn
\chi_{zj}^{(e)} = (U_x(\pi))_{zz}\chi_{zj}^{(e)} = -\chi_{zj}^{(e)},
\eeqn
which forces $\chi_{zj}^{(e)}=0$. Consequently, all components vanish:
\beqn
\chi_{\alpha j}^{(e)} = 0, \qquad \forall \alpha,j.
\eeqn
Thus, the $T$-even Edelstein effect is strictly forbidden in altermagnets.

\subsection{H2. Proof of the absence of $T$-odd spin current in odd‑parity magnets}

Odd-parity magnets are defined by the condition $\bm{S}(\bm{k}) = -\bm{S}(-\bm{k})$ for the spin texture. The symmetry criteria for odd-parity magnets have been established in Ref.~\cite{2026Luoxun}. Here we show that the $T$-odd spin-current conductivity $\sigma_{ij}^{\alpha(o)}$ vanishes in odd-parity magnets.

For a $T$-odd tensor, the transformation under   a SSG operation $\{\mathcal{U}\|R|\bm \tau\}$  is
\beqn
\sigma_{ij}^{\alpha(o)} = \sum_{\beta,m,n} \mathcal{U}_{\alpha\beta} R_{im} R_{jn} \sigma_{mn}^{\beta(o)}. \label{ss}
\eeqn
For nonpolanar magnetic orders exhibiting odd-parity spin splitting, the  symmetry  $\{T \| I|\bm{\tau}\}$ is needed \cite{2026Luoxun}. This symmetry enforces
\beqn
\sigma_{ij}^{\alpha(o)} = -\sigma_{ij}^{\alpha(o)}=0.
\eeqn
Odd‑parity magnets with coplanar magnetic orders (e.g., with spins lying in the $xy$ plane) typically possess two symmetries: 
$\{U_z(\theta) \| I|\bm 0\}$ and $\{T U_z(\pi) \| I|\bm 0\}$. For $\{U_z(\theta) \| I|\bm 0\}$, Eq.~\ref{ss} applies with $\mathcal{U} = U_z(\theta)$ and $R = I$. For each fixed spatial pair $(i,j)$, the three components $(\sigma_{ij}^{x}, \sigma_{ij}^{y}, \sigma_{ij}^{z})$ form a vector that must be invariant under $U_z(\theta)$ for all $\theta$. The only invariant direction is the $z$ axis, hence
$\sigma_{ij}^{x(o)} = \sigma_{ij}^{y(o)} = 0$.
Thus, $\{U_z(\theta) \| I|\bm 0\}$ forces the spin current conductivity to have only the $z$ spin component. The spin operation of symmetry $\{T U_z(\pi) \| I|\bm 0\}$ is $ \operatorname{diag}(1,1,-1)$ in the spin space, which enforces $\sigma_{ij}^{z(o)}=0$ .  Hence, odd‑parity magnets with coplanar magnetic orders  exhibit no $T$-odd spin current.

\section{I. Type-II and type-III hybrid-parity magnets}

Type-II hybrid-parity magnets are characterized by a noncoplanar spin texture with two even-parity components and one odd-parity component, i.e., $s_{x,z}(\bm{k}) = s_{x,z}(-\bm{k})$ and $s_y(\bm{k}) = -s_y(-\bm{k})$, enforced by the symmetry $\{T U_y(\pi) \| I|\bm{0}\}$ alone. This symmetry imposes the following constraints on the response tensors:
\beqn
\sigma_{ij}^{y(o)} = 0, \qquad \chi_{xj}^{(e)} = \chi_{zj}^{(e)} = 0,
\eeqn
while $\sigma_{ij}^{x(o)}$, $\sigma_{ij}^{z(o)}$, and $\chi_{yj}^{(e)}$ remain symmetry-allowed. These constraints reflect the parity of the spin texture: the even-parity components $s_{x,z}$ indicate $T$-odd spin currents polarized along $x$ and $z$, whereas the odd-parity component $s_y$ indicates the $T$-even Edelstein effect polarized along $y$. This behavior illustrates the intimate relation between spin-texture parity and Fermi-surface spin transport.

Type-III hybrid-parity magnets are characterized by a noncoplanar spin texture with two odd-parity components and one even-parity component, i.e., $s_{x,y}(\bm{k}) = -s_{x,y}(-\bm{k})$ and $s_z(\bm{k}) = s_z(-\bm{k})$, enforced by the symmetry $\{U_z(\pi) \| P|\bm{0}\}$ alone. This symmetry imposes
\beqn
\sigma_{ij}^{x(o)} = \sigma_{ij}^{y(o)} = 0, \qquad \chi_{zj}^{(e)} = 0,
\eeqn
while $\sigma_{ij}^{z(o)}$, $\chi_{xj}^{(e)}$, and $\chi_{yj}^{(e)}$ remain symmetry-allowed. These constraints also reflect the parity of the spin texture: the even-parity component $s_z$ indicates the $T$-odd spin current polarized along $z$, whereas the odd-parity components $s_{x,y}$ indicate the $T$-even Edelstein effect polarized along $x$ and $y$.

From the MAGNDATA database, we identify one candidate material, FeSb$_2$O$_4$, described by the oriented SSG $P^{2_{010}}4_2/^{2_{001}}m^{2_{100}}b^{2_{001}}c$. Its symmetry generators are $\{U_x(\pi)\|M_x|\bm{\tau}_{(1/2,1/2,0)}\}$, $\{U_z(\pi)\|M_z|\bm{0}\}$, $\{U_y(\pi)\|R_{4z}|\bm{\tau}_{(0,0,1/2)}\}$, and $\{U_x(\pi)\|M_y|\bm{\tau}_{(1/2,1/2,0)}\}$. These symmetries impose the following constraints on the spin textures:
\begin{align}
\{U_x(\pi)\|M_x|\bm{\tau}_{(1/2,1/2,0)}\}:&\quad s_x(k_x,k_y,k_z) = s_x(-k_x,k_y,k_z),\quad s_{y,z}(k_x,k_y,k_z) = -s_{y,z}(-k_x,k_y,k_z),\nonumber\\
\{U_z(\pi)\|M_z|\bm{0}\}:&\quad s_{x,y}(k_x,k_y,k_z) = -s_{x,y}(k_x,k_y,-k_z),\quad s_z(k_x,k_y,k_z) = s_z(k_x,k_y,-k_z),\nonumber\\
\{U_y(\pi)\|R_{4z}|\bm{\tau}_{(0,0,1/2)}\}:&\quad s_{x,z}(k_x,k_y,k_z) = -s_{x,z}(k_y,-k_x,k_z),\quad s_y(k_x,k_y,k_z) = s_y(k_y,-k_x,k_z),\nonumber\\
\{U_x(\pi)\|M_y|\bm{\tau}_{(1/2,1/2,0)}\}:&\quad
s_{x}(k_x,k_y,k_z) = s_{x}(k_x,-k_y,k_z),\quad s_{y,z}(k_x,k_y,k_z) = -s_{y,z}(k_x,-k_y,k_z).
\end{align}
From these constraints, the spin textures $(s_x,s_y,s_z)$ form a reducible three-dimensional representation $B_{2u}\oplus A_{1u}\oplus B_{2g}$ of the point group $D_{4h}$ ($4/mmm$), characterized by
\begin{equation}
s_x(\bm{k})\propto (k_x^2-k_y^2)k_z,\quad
s_y(\bm{k})\propto k_x k_y(k_x^2-k_y^2)k_z,\quad
s_z(\bm{k})\propto k_x k_y.
\end{equation}
Therefore, the components $s_x$, $s_y$, and $s_z$ exhibit $f$-, $h$-, and $d$-wave characters, respectively.

Under the constraints of these symmetries, the response tensors take the form
\begin{equation}
\chi_{\alpha j}^{(e)}=0,\quad \sigma_{ij}^{x(o)}=\sigma_{ij}^{y(o)}=0,\quad
\sigma^{z(o)} = \begin{pmatrix}
0 & \sigma_{xy}^{z} & 0 \\
\sigma_{yx}^{z} & 0 & 0 \\
0 & 0 & 0
\end{pmatrix}.
\end{equation}
This example demonstrates that the presence of odd-parity spin textures is a necessary but not sufficient condition for the Edelstein effect: additional symmetries that do not constrain the parity of the spin textures may still enforce the vanishing of the response tensor. The same applies to the $T$-odd spin current.

For type-II HPMs, we identify 17 materials. Taking TbFeO$_3$ as an example, it is described by the oriented SSG $P^{2_{001}}2_1^{2_{010}}2_1^{2_{100}}2_1^{m_{100}}1$. The symmetry generators are 
\[
\{U_z(\pi)\|R_{x}(\pi)|\bm{\tau}_{(\frac{1}{2},\frac{1}{2},0)}\},\quad
\{U_y(\pi)\|R_y(\pi)|\bm{\tau}_{(0,\frac{1}{2},\frac{1}{2})}\},\quad
\{U_x(\pi)\|R_z(\pi)|\bm{\tau}_{(\frac{1}{2},0,\frac{1}{2})}\},\quad
\{T U_x(\pi)\|I|\bm{0}\}.
\]
These symmetries impose the following constraints on the spin texture $\bm{S}(\bm{k})$:
\begin{align}
\{U_z(\pi)\|R_{x}(\pi)|\bm{\tau}_{(\frac{1}{2},\frac{1}{2},0)}\}&:\quad 
s_{x,y}(k_x,k_y,k_z) = -s_{x,y}(k_x,-k_y,-k_z),\; s_{z}(k_x,k_y,k_z) = s_{z}(k_x,-k_y,-k_z),\nonumber\\
\{U_y(\pi)\|R_y(\pi)|\bm{\tau}_{(0,\frac{1}{2},\frac{1}{2})}\}&:\quad 
s_{x,z}(k_x,k_y,k_z) = -s_{x,z}(-k_x,k_y,-k_z),\; s_{y}(k_x,k_y,k_z) = s_{y}(-k_x,k_y,-k_z),\nonumber\\
\{U_x(\pi)\|R_z(\pi)|\bm{\tau}_{(\frac{1}{2},0,\frac{1}{2})}\}&:\quad 
s_{y,z}(k_x,k_y,k_z) = -s_{y,z}(-k_x,-k_y,k_z),\; s_{x}(k_x,k_y,k_z) = s_{x}(-k_x,-k_y,k_z),\nonumber\\
\{T U_x(\pi)\|I|\bm{0}\}&:\quad 
s_{y,z}(\bm{k}) = s_{y,z}(-\bm{k}),\; s_{x}(\bm{k}) = -s_{x}(-\bm{k}).
\end{align}
From these constraints, we find that $(s_x(\bm{k}), s_y(\bm{k}), s_z(\bm{k}))$ form a reducible three-dimensional representation $B_{1u} \oplus B_{2g} \oplus B_{3g}$ of the point group $D_{2h}$ ($mmm$), characterized by the lowest-order expansions
\beqn
s_x(\bm{k}) \propto k_z,\quad s_y(\bm{k}) \propto k_x k_z,\quad s_z(\bm{k}) \propto k_y k_z.
\eeqn

Under these symmetry constraints, the response tensors take the form
\begin{align}
\sigma^{x(o)} &= \begin{pmatrix}
0 & 0 & 0 \\
0 & 0 & 0 \\
0 & 0 & 0
\end{pmatrix},\quad
\sigma^{y(o)} = \begin{pmatrix}
0 & 0 & \sigma_{xz}^{y} \\
0 & 0 & 0 \\
\sigma_{zx}^{y} & 0 & 0
\end{pmatrix},\quad
\sigma^{z(o)} = \begin{pmatrix}
0 & 0 & 0 \\
0 & 0 & \sigma_{yz}^{z} \\
0 & \sigma_{zy}^{z} & 0
\end{pmatrix},\nonumber\\
\chi^{(e)} &= \begin{pmatrix}
0 & 0 & \chi_{xz} \\
0 & 0 & 0 \\
0 & 0 & 0
\end{pmatrix}.
\end{align}
Thus, type-II HPMs can generate spin currents with spin polarizations along two distinct directions ($y$ and $z$ in TbFeO$_3$), and can also exhibit an Edelstein response. This multi-component transport behavior is a distinctive feature of type-II HPMs.

	\begin{figure}[htbp]
		\begin{centering}
			\includegraphics[width=\textwidth]{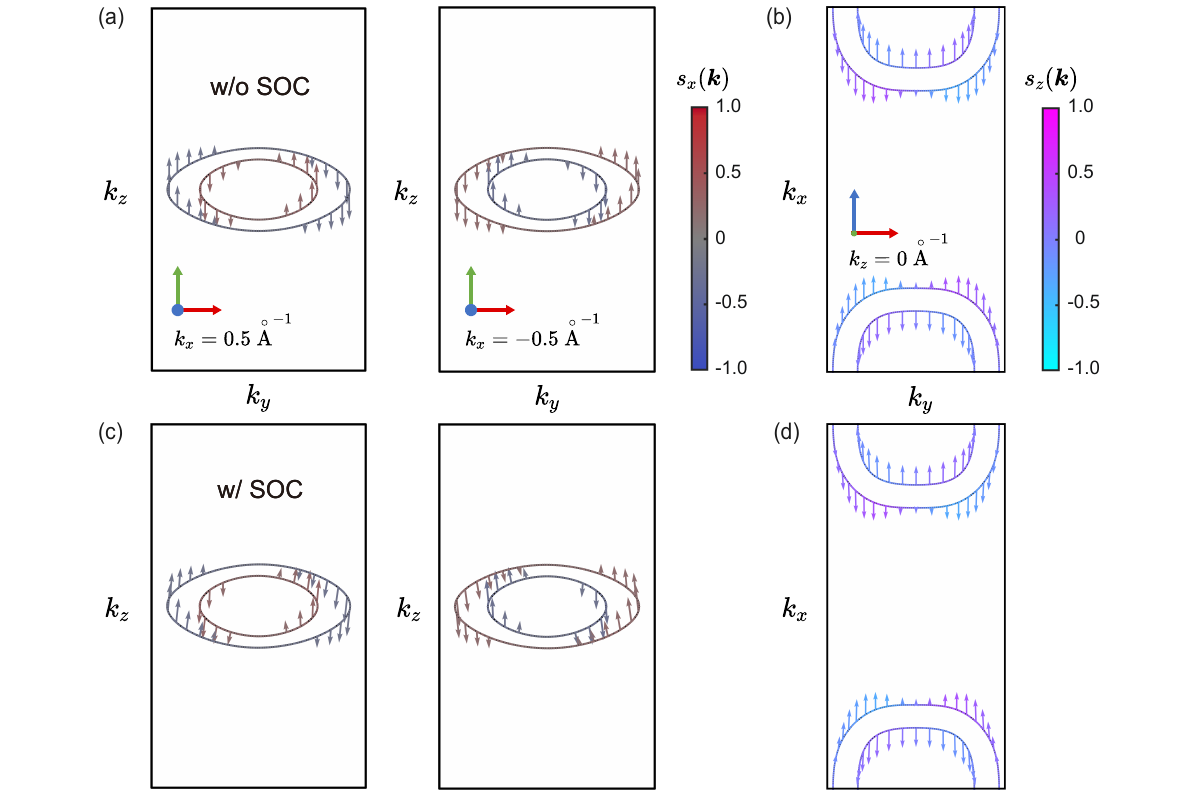}
			\par\end{centering}
		\caption{
		Calculated spin textures of \ch{FePO4} (a, b) without and (c, d) with SOC. Panels (a) and (c) display the spin textures in the $k_y$-$k_z$ planes located at $k_x = \pm 0.5~\text{\AA}^{-1}$. Panels (b) and (d) show the spin textures in the $k_x$-$k_y$ plane at $k_z = 0~\text{\AA}^{-1}$. In all panels, the local direction of the arrows represents the in-plane spin expectation values, while the color mapping encodes the magnitude of the out-of-plane spin component, namely $s_x(\boldsymbol{k})$ in (a) and (c), and $s_z(\boldsymbol{k})$ in (b) and (d).
		}
		\label{fig:figS3} 
	\end{figure}

	\begin{figure}[htbp]
		\begin{centering}
			\includegraphics[width=\textwidth]{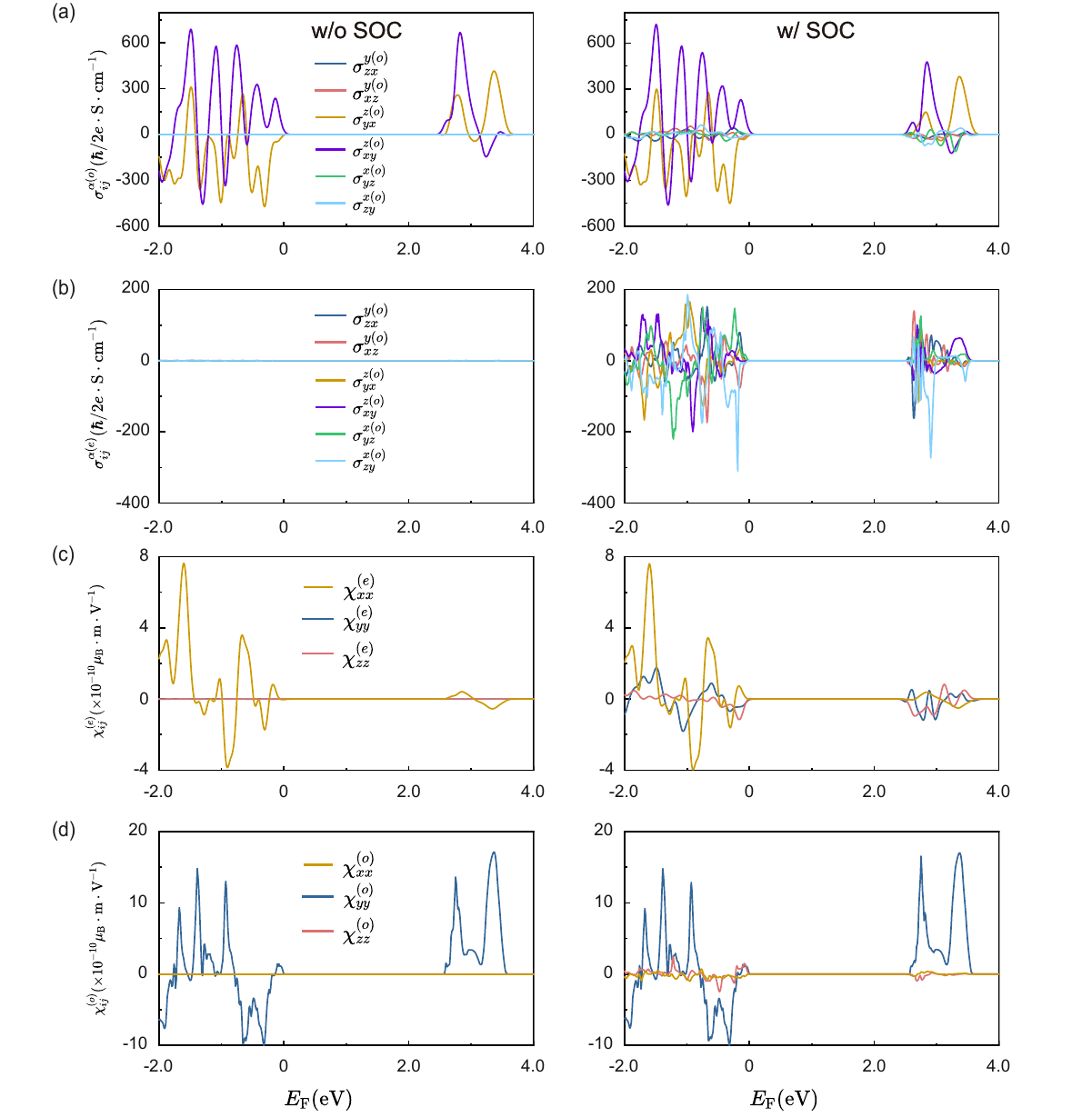}
			\par\end{centering}
		\caption{
		Calculated Fermi-energy $E_{\text{F}}$ dependence of various spin transport response tensors in \ch{FePO4}. The panels display (a, b) the spin current conductivities $\sigma^{\alpha(o)}_{ij}$ and $\sigma^{\alpha(e)}_{ij}$, alongside (c, d) the Edelstein susceptibilities $\chi^{(e)}_{ij}$ and $\chi^{(o)}_{ij}$. In each row, the left panel represents the calculation results neglecting SOC, while the right panel includes the effect of SOC. Evaluated tensor components are distinguished by different colors as indicated in the respective legends.
		}
		\label{fig:figS4} 
	\end{figure}

\end{widetext}

\end{document}